%% file: main.tex
\documentclass[twocolumn,amsmath,amssymb,aps,pra,superscriptaddress]{revtex4-2}

\usepackage[T1]{fontenc}
\usepackage[utf8]{inputenc}
\usepackage[english]{babel}
\usepackage{amsmath}
\usepackage{amssymb}
\usepackage{graphicx}
\usepackage{dcolumn}
\usepackage{siunitx}
\usepackage{physics}
\usepackage{xcolor}
\usepackage[colorlinks=true,linkcolor=blue!60!black,citecolor=red!60!black,urlcolor=blue!60!black]{hyperref}

\renewcommand{\vec}[1]{\mbox{\boldmath ${#1}$}}

\begin{document}

\title{Dynamics of Cavity-Induced Radiative Stabilization of a Dissociating Polarized Molecular Ion}

\author{Martin Mendez}
\email{martinmendez@unc.edu.ar}
\affiliation{Facultad de Matem\'atica, Astronom\'ia, F\'isica y Computaci\'on, Universidad Nacional de C\'ordoba, and Instituto de F\'isica Enrique Gaviola, CONICET-UNC, Ciudad Universitaria, C\'ordoba X5000HUA, Argentina}

\author{Federico M. Pont}
\email{pont.federico@unc.edu.ar}
\affiliation{Facultad de Matem\'atica, Astronom\'ia, F\'isica y Computaci\'on, Universidad Nacional de C\'ordoba, and Instituto de F\'isica Enrique Gaviola, CONICET-UNC, Ciudad Universitaria, C\'ordoba X5000HUA, Argentina}

\date{\today}

\input{section_abstract}

\maketitle

\input{section_introduction}
\input{section_model}
\input{section_methods}
\input{section_results_small}
\input{section_discussion}
\input{section_conclusions}
\input{section_acknowledgments}

\appendix
\input{appendix_hamiltonians}
\input{appendix_trk}
\input{appendix_photon}

\bibliography{bibliography}

\end{document}

%% file: section_abstract.tex
\begin{abstract}
Models of cavity-modified molecular dynamics often neglect the permanent dipoles of the molecule and the dipole self-energy (DSE) of the Pauli--Fierz Hamiltonian. We show that both matter once the nuclei move. In a one-dimensional model of the NeHe$^+$ ion excited to a dissociative state, the cavity radiatively stabilizes the molecule: at the resonant avoided crossing, the molecule emits a photon into the cavity and is trapped in its bound ground state. The trapped fraction rises from $14\%$ to $49\%$ as the passage through the crossing turns from diabatic to adiabatic, while stronger couplings populate electronic states beyond the two-level manifold. A full electron--nucleus--photon simulation confirms the stabilization and shows that it entangles the photon number with the fate of the molecule, trapped or dissociating. The length-gauge photonic observables are dominated by the static polarization field of the permanent dipoles, whereas the free photons consist essentially of the single emitted photon and have sub-Poissonian statistics.
\end{abstract}

\keywords{cavity quantum electrodynamics, Pauli--Fierz Hamiltonian, dipole self-energy, polaritonic chemistry, molecular dissociation, MCTDH}

%% file: section_introduction.tex
\section{Introduction}
\label{sec:introduction}

The interaction of atoms and molecules with confined electromagnetic fields has become a central topic in quantum optics and chemical physics. Cavity quantum electrodynamics (cQED) provides a platform in which the light--matter coupling modifies the energetic, dynamical, and chemical properties of microscopic systems. Strong coupling between single atoms and quantized cavity fields was first demonstrated in Fabry--P\'erot resonators~\cite{Thompson1992}, opening the way to the coherent manipulation of individual atoms and photons~\cite{haroche2013nobel}. In molecular systems, the hybrid light--matter eigenstates (polaritons), which can often be approximated as bosonic quasiparticles although their effective statistics depends on their constituents and on the excitation density~\cite{Ruggenthaler2018,ruggenthaler_understanding_2023}, have been shown to reshape potential-energy surfaces, to alter chemical reaction rates, and to generate nonclassical states of light~\cite{Ebbesen2016,Feist2018,Ruggenthaler2018,GarciaVidal2021}. More generally, light--matter interactions that are nonlinear in the photon number can turn a coherent field into nonclassical states~\cite{Uria2020PRL,Uria2023PRR}. Experiments demonstrated that collective strong coupling inside a Fabry--P\'erot cavity can slow down a photoisomerization reaction~\cite{Hutchison2012}, and theory showed that the coupling can suppress photochemical reactions by modifying the excited-state landscape~\cite{Galego2016}. In the single-molecule limit, strong coupling has been reached at room temperature in plasmonic nanogaps~\cite{Chikkaraddy2016}, where atomic-scale features confine light to volumes below $1\ \text{nm}^3$~\cite{Benz2016}.

The paradigmatic models of quantum optics---Jaynes--Cummings, Rabi, and Dicke---describe limits of the light--matter interaction that can be derived from the nonrelativistic Pauli--Fierz (PF) Hamiltonian~\cite{ruggenthaler_understanding_2023,Taylor2024}. The Rabi model retains the counter-rotating terms that become relevant in the ultrastrong-coupling (USC) regime~\cite{forn2019ultrastrong,frisk2019ultrastrong}, and the Dicke model describes an ensemble of emitters whose superradiant transition in the thermodynamic limit~\cite{Dicke1954,Hepp1973} is forbidden when the diamagnetic or dipole self-energy (DSE) term is retained~\cite{Rzazewski1975,Bialynicki1979}. In the length gauge, the DSE is the term that makes the Hamiltonian bounded from below and gauge consistent~\cite{Rokaj2018,schafer_relevance_2020,DiStefano2019,Taylor2020}; its truncation to a finite electronic basis is, however, not unique~\cite{DiStefano2019,Taylor2020,Ryu2023,Weight2023,Taylor2024}, and whether it is required at all in realistic nanophotonic setups, where the coupling is dominated by quasistatic Coulomb interactions (i.e., by the near field of the charges of the nanostructure rather than by transverse photons), has recently been questioned~\cite{Pradilla2025}.

Molecules add new layers of complexity with respect to atoms: coordinate-dependent transition energies and dipole moments, permanent dipole moments (PDMs) that grow along a dissociation coordinate in heteronuclear species, and light-induced avoided crossings and conical intersections that reshape nonadiabatic pathways. \textit{Ab initio} approaches to these problems include quantum-electrodynamical density-functional theory~\cite{Flick2017}, the cavity Born--Oppenheimer approximation~\cite{Flick2017cBO}, the exact factorization~\cite{Lacombe2019,Abedi2018}, polaritonic potential-energy surfaces in the electronic strong-coupling regime~\cite{Galego2015,Galego2016}, exact simulations of small molecules~\cite{Sidler2020,Krupp2026}, and the multiconfiguration time-dependent Hartree (MCTDH) method~\cite{Meyer1990,Beck2000,MeyerMCTDHpackage}, which has been applied to collective Jahn--Teller physics~\cite{vendrell_collective_2018}, coherent femtochemistry~\cite{vendrell_coherent_2018}, and cavity control of molecular dynamics~\cite{csehi_controlling_2025}. Two different partitions of the coupled electron--nucleus--photon problem are used in the literature~\cite{ruggenthaler_understanding_2023}: the cavity Born--Oppenheimer partition, which groups the photon displacement coordinate with the nuclei and is natural when the cavity is resonant with vibrations~\cite{Flick2017cBO}, and the polaritonic-surface partition, which groups photons with electrons and is natural when the cavity is resonant with an electronic transition~\cite{Galego2015}. Most treatments either fix the nuclear geometry, retain only two electronic states, or neglect the PDMs and the DSE, so that the combined effect of these ingredients on a dissociative nuclear wave packet has remained largely unexplored.

Interparticle Coulombic decay (ICD)~\cite{Cederbaum1997,Jahnke2020} and interparticle Coulombic electron capture (ICEC)~\cite{Gokhberg2009,Gokhberg2010} are environment-assisted processes in which energy is transferred between neighboring atoms or molecules through long-range electron correlation. In ICD, an excited species relaxes by ionizing a neighbor; in ICEC, a free electron is captured by one species and the released energy ionizes a neighbor, a mechanism that has also been studied in quantum dots~\cite{Pont2016electron,Pont2020predicting}. In the near-field regime relevant for these processes, the energy is carried by a virtual photon, and, when these channels are open, ICD is orders of magnitude faster than radiative decay and ICEC can dominate over photorecombination up to nanometer distances~\cite{Jahnke2020,Gokhberg2010}. A cavity concentrates the coupling to the electromagnetic field, which in free space is spread over a continuum of modes, into one or a few discrete modes with a large coupling strength. These modes become dynamical degrees of freedom of the system, and their occupation by real photons opens energy-transfer channels that coexist with the near-field Coulomb exchange and can be tuned through the cavity frequency and mode volume: cavity polaritons can enhance or suppress ICD~\cite{Cederbaum2021}, and a cavity photon can transfer energy from an excited atom to a molecule and trigger its dynamics~\cite{csehi_controlling_2025}. Nuclear motion can also change these processes qualitatively. In the NeHe$^+$ molecular ion, ICEC occurs below the threshold imposed by the fixed-nuclei approximation, and the correlations generated during the capture have been interpreted in terms of the potential-energy curves (PECs) of the molecule and of the transitions between them~\cite{Pont2024impact,Mendez2025}. Extending this analysis to a molecule inside a cavity requires a description of the cavity field and a scheme of dressed states, analogous to the Born--Oppenheimer PECs, in which capture, decay, and dissociation are transitions between light--matter states. The present work constructs and validates such a scheme for the simplest process in which electronic, nuclear, and photonic degrees of freedom all take part: the dissociation of the NeHe$^+$ ion excited inside a cavity.

In this work, we study the coupled nucleus--photon dynamics of a one-dimensional model of NeHe$^+$ in a single cavity mode resonant with the vertical excitation at the equilibrium distance. The model, introduced in previous work on electron capture~\cite{Mendez2025,Pont2024impact}, describes the electronic continuum through pseudocontinuum states computed on a finite-element grid, and its reduced dimensionality allows us to compare a hierarchy of approximations with a full electron--nucleus--photon propagation (Fig.~\ref{fig:models}). Each level of the hierarchy defines its own potential-energy curves (molecular or polaritonic), which provide a physical picture of the dynamics in terms of transitions between them and show where the simpler descriptions fail. Explicitly, we (i) retain ten adiabatic electronic states with their permanent and transition dipoles and a DSE matrix; (ii) show that, for the truncated multistate Hamiltonian, the DSE coefficient prescribed by the PF Hamiltonian is necessary and sufficient for the invariance of the spectrum under a change of dipole origin in both truncated forms of the DSE, extending the known gauge-consistency argument~\cite{Rokaj2018,schafer_relevance_2020,Taylor2020}, and bound the difference between the two forms with the Thomas--Reiche--Kuhn (TRK) sum rule; (iii) validate the polaritonic PECs against a diagonalization in which electrons and photons are treated simultaneously without electronic truncation; (iv) propagate the nucleus--photon wave packet for eight coupling strengths; and (v) validate this dynamics, for the largest coupling, against the full electron--nucleus--photon propagation. The cavity opens a radiative stabilization channel: at the resonance, the excited molecule emits a photon into the cavity and is transferred to its bound electronic ground state, where it can no longer dissociate. This channel partially quenches the dissociation, with a yield controlled by the crossover from diabatic to adiabatic passage. The length-gauge field coherence is quantitatively accounted for by the static polarization generated by the PDMs, whereas the free photons, their statistics, the light--matter entanglement, and the breakdown of the two-level approximation are characterized as functions of the coupling.

The paper is organized as follows. Section~\ref{sec:model} introduces the molecular model, the PF Hamiltonian, its multistate reduction, the two forms of the DSE, and the hierarchy of approximate models. Section~\ref{sec:methods} describes the electronic-structure, polaritonic, and dynamical methods and the observables. Section~\ref{sec:results} presents the static and dynamical results, Sec.~\ref{sec:discussion} discusses them in the context of previous work and of experimentally accessible cavities, and Sec.~\ref{sec:conclusions} summarizes. The Hamiltonians of the model hierarchy, the Thomas--Reiche--Kuhn bound on the DSE gap, and the relation between length-gauge and free-photon observables are detailed in the Appendices, and the proof of gauge invariance, the validation of the polaritonic PECs, the numerical details and convergence tests, and the two-level models without PDMs are given in the Supplemental Material~\cite{SM}\nocite{HornJohnson2012,ColbertMiller1992,Riss1993,MeyerWang2018}. Atomic units are used throughout unless stated otherwise.

%% file: section_model.tex
\section{Model}
\label{sec:model}

\subsection{Molecular model}
\label{sec:molecular_model}

The molecular model was introduced in Refs.~\cite{Mendez2025,Pont2024impact} to describe electron capture by NeHe$^+$. One active electron interacts with two nuclei, Ne and He, through Yukawa-type potentials with effective atomic numbers $\mathcal{Z}_{\text{Ne}}=1.3068$ and $\mathcal{Z}_{\text{He}}=1.4532$, and a transverse harmonic confinement reduces the problem to a single electronic coordinate $z$ along the internuclear axis and the internuclear distance $R$. The model is therefore a molecular \emph{ion}: its net charge, $(\mathcal{Z}_{\text{Ne}}+\mathcal{Z}_{\text{He}}-1)|e|$, does not vanish, and its dipole moment depends on the choice of origin (Sec.~\ref{sec:dse}). The two lowest PECs are energetically separated from the rest; the first excited PEC is dissociative in the Franck--Condon region, and the states correlate asymptotically with the electron localized on different centers, which gives rise to charge-transfer PDMs that grow linearly with $R$.

The choice of this system has three motivations. First, its heteronuclear character produces large, $R$-dependent PDMs. Second, the dissociative first excited state provides a clean setting to study whether the cavity can stabilize a molecule against dissociation after a vertical excitation. Third, the one-dimensional single-active-electron description gives access, on the same finite-element grid, to the bound states and to a discretized continuum, and hence to the exact matrix elements of $\hat{\mu}^2$ and to a reference calculation in which electrons and photons are diagonalized simultaneously without electronic truncation (Sec.~\ref{sec:methods_pecs}). These quantities cannot be obtained with the same control from standard quantum-chemistry methods, whose Gaussian basis sets are designed for bound states and represent the continuum only through specially optimized diffuse functions~\cite{Kaufmann1989,Coccia2016}, and they are central to the analysis of the DSE.

The molecule is assumed to be oriented, with its axis parallel to the polarization $\vec{\epsilon}_0$ of a single cavity mode of frequency $\omega_0$ resonant with the vertical electronic excitation at the equilibrium distance. The dipole approximation is adopted, and cavity losses are neglected; their role is discussed in Sec.~\ref{sec:disc_feasibility}.

\subsection{Pauli--Fierz Hamiltonian in the length gauge}
\label{sec:pf_hamiltonian}

The starting point is the nonrelativistic PF Hamiltonian in the dipole approximation~\cite{ruggenthaler_understanding_2023,Taylor2024}. The molecule is placed at a fixed point $\vec{r}_0$ of the cavity, the field is taken as uniform over the molecule, and its translational motion is neglected. The Power--Zienau--Woolley (PZW) transformation $\hat{U}_{\text{PZW}}=\exp[-\mathrm{i}\,\hat{\vec{\mu}}\cdot\hat{\vec{A}}(\vec{r}_0)/\hbar]$~\cite{PowerZienau1959,Woolley1971}, where $\hat{\vec{\mu}}$ is the total (electronic plus nuclear) electric dipole operator of the molecule and $\hat{\vec{A}}(\vec{r}_0)$ is the vector-potential operator of the cavity mode evaluated at $\vec{r}_0$, followed by a phase rotation of the mode, brings the Coulomb-gauge Hamiltonian into the length-gauge form
\begin{equation}\label{eq:PF_ham}
  \hat{H}_{\text{PF}} = \hat{H}_{\text{mol}} + \hbar\omega_0\Big(\hat{a}^{\dagger}\hat{a}+\tfrac12\Big) + g\,(\hat{a}+\hat{a}^{\dagger})\,\hat{\mu}_\epsilon + \frac{g^2}{\hbar\omega_0}\,\hat{\mu}_\epsilon^{\,2},
\end{equation}
where $\hat{H}_{\text{mol}}$ is the molecular Hamiltonian, $\hat{a}$ and $\hat{a}^{\dagger}$ are the bosonic annihilation and creation operators of the cavity mode, $[\hat{a},\hat{a}^{\dagger}]=1$, $\hat{\mu}_\epsilon\equiv\vec{\epsilon}_0\cdot\hat{\vec{\mu}}$ is the projection of the dipole operator on the polarization, and

\begin{equation}\label{eq:g_def}
  g=\lambda\sqrt{\frac{\hbar\omega_0}{2}}
\end{equation}
is the vacuum-field amplitude of a mode of effective volume $V$, with $\lambda=1/\sqrt{\varepsilon_0V}$ the coupling strength used in \textit{ab initio} QED~\cite{ruggenthaler_understanding_2023,Flick2017}. The number states $|n\rangle$ of the mode ($\hat{a}^{\dagger}\hat{a}|n\rangle=n|n\rangle$) form the Fock basis used below; after the PZW transformation, $\hat{a}$ and $\hat{a}^{\dagger}$ no longer annihilate and create free photons (Eq.~\eqref{eq:physical_photon}).

A different definition of the coupling constant is common in the literature. In quantum optics and in many models of molecular polaritons, the dipole moment is included in the coupling, which is then the vacuum Rabi frequency $g\,\mu_{12}/\hbar$ of a given transition~\cite{jaynes1963comparison,haroche2013nobel,Kowalewski2016,Feist2018}. This convention is convenient when the dipole moment is a fixed number, but here the dipole matrix elements depend on the internuclear distance, which is a dynamical variable, so that a coupling that includes them is no longer a constant. The definition of Eq.~\eqref{eq:g_def}, which depends only on the cavity, is therefore the only one used in this work: the coupling between the electronic states $i$ and $j$ is $g\,\mu^{\text{BOA}}_{ij}(R)$, with the dipole matrix elements defined below. The last term of Eq.~\eqref{eq:PF_ham} is the DSE.

Using $\hat{Q}=\sqrt{\hbar/2\omega_0}\,(\hat{a}+\hat{a}^{\dagger})$ and $\hat{P}=\mathrm{i}\sqrt{\hbar\omega_0/2}\,(\hat{a}^{\dagger}-\hat{a})$, the field part of Eq.~\eqref{eq:PF_ham} is~\cite{Rokaj2018},
\begin{equation}\label{eq:completed_square}
  \begin{split}
  &\hbar\omega_0\Big(\hat{a}^{\dagger}\hat{a}+\tfrac12\Big) + g(\hat{a}+\hat{a}^{\dagger})\hat{\mu}_\epsilon + \frac{g^2}{\hbar\omega_0}\hat{\mu}_\epsilon^{\,2}\\
  &\qquad= \frac{\hat{P}^2}{2}+\frac{\omega_0^2}{2}\Big(\hat{Q}+\frac{\lambda\hat{\mu}_\epsilon}{\omega_0}\Big)^{2}.
  \end{split}
\end{equation}
In the Coulomb gauge, the momentum conjugate to the vector potential is proportional to the transverse electric field. The PZW transformation leaves the vector potential unchanged but redefines its conjugate momentum, which becomes proportional to the electric displacement field; the phase rotation, which replaces $\hat{a}$ by $\mathrm{i}\hat{a}$ and thus $\hat{Q}$ by $-\hat{P}/\omega_0$ and $\hat{P}$ by $\omega_0\hat{Q}$ so that the coupling of Eq.~\eqref{eq:PF_ham} is real, exchanges the roles of coordinate and momentum, and the mode coordinate represents the displacement field~\cite{CohenTannoudji1989,CraigThirunamachandran1984,Abedi2018,schafer_relevance_2020}. Let $\hat{D}$, $\hat{E}_\perp$, and $\hat{P}_{\text{pol}}$ denote the components along $\vec{\epsilon}_0$ of the electric displacement-field operator, of the transverse electric-field operator, and of the transverse polarization operator of the molecule, respectively, which are related by $\hat{D}=\varepsilon_0\hat{E}_\perp+\hat{P}_{\text{pol}}$~\cite{CohenTannoudji1989,CraigThirunamachandran1984}; in the dipole approximation $\hat{P}_{\text{pol}}$ is determined by the molecular dipole operator $\hat{\mu}_\epsilon$. Then $\omega_0\hat{Q}$ and $\omega_0\hat{Q}+\lambda\hat{\mu}_\epsilon$ are proportional, with a common negative constant, to $\hat{D}$ and $\varepsilon_0\hat{E}_\perp$, respectively, so that their difference, $\hat{P}_{\text{pol}}$, is proportional to $+\hat{\mu}_\epsilon$. Accordingly, the operator that annihilates a free (transverse) photon is not $\hat{a}$, but its PZW image $\hat{a}_{\text{PZW}}$~\cite{schafer_relevance_2020,Taylor2024,Welakuh2025multimode}, i.e., the photon annihilation operator dressed by the polarization of the medium due to the molecule,
\begin{equation}\label{eq:physical_photon}
  \begin{split}
  \hat{a}_{\text{PZW}}&=\hat{U}_{\text{PZW}}\,\hat{a}\,\hat{U}_{\text{PZW}}^\dagger=\hat{a}+\frac{g}{\hbar\omega_0}\,\hat{\mu}_\epsilon,\\
  \hat{n}_{\text{PZW}}&\equiv\hat{a}_{\text{PZW}}^\dagger\hat{a}_{\text{PZW}}\\
  &=\hat{n}+\frac{g}{\hbar\omega_0}(\hat{a}+\hat{a}^{\dagger})\hat{\mu}_\epsilon+\frac{g^2}{\hbar^2\omega_0^2}\hat{\mu}_\epsilon^{\,2},
  \end{split}
\end{equation}
where $\hat{n}\equiv\hat{a}^{\dagger}\hat{a}$ is the length-gauge photon-number operator and, written with the rotated mode operators, $\hat{U}_{\text{PZW}}=\exp[-(g/\hbar\omega_0)\hat{\mu}_\epsilon(\hat{a}^{\dagger}-\hat{a})]$ is a displacement of the mode by the dipole-dependent amplitude $-(g/\hbar\omega_0)\hat{\mu}_\epsilon$, which commutes with $\hat{a}$. Thus $\hat{a}_{\text{PZW}}$ is the Coulomb-gauge photon operator expressed in the length gauge (up to the constant phase of the rotation), and $\hat{U}_{\text{PZW}}^\dagger$ maps a length-gauge state back to the Coulomb gauge. The field part of Eq.~\eqref{eq:PF_ham} is simply $\hbar\omega_0(\hat{n}_{\text{PZW}}+\tfrac12)$. Expectation values of $\hat{a}$ and $\hat{n}$ are therefore length-gauge quantities that contain, besides free photons, the static field bound to the molecular polarization. This distinction is essential for the interpretation of the photonic observables in Sec.~\ref{sec:results}.

The molecular Hamiltonian is $\hat{H}_{\text{mol}}=\hat{T}_R+\hat{H}_{\text{el}}(z;R)$, with $\hat{T}_R=-(\hbar^2/2\mathcal{M}_{\text{red}})\partial_R^2$ the nuclear kinetic energy, $\mathcal{M}_{\text{red}}$ the nuclear reduced mass, and $\hat{H}_{\text{el}}$ the electronic Hamiltonian at fixed $R$~\cite{Mendez2025,Pont2024impact}. Taking the origin at the nuclear center of mass, the nuclei are located at $z_{\text{Ne}}=-\mathcal{M}_{\text{He}}R/(\mathcal{M}_{\text{Ne}}+\mathcal{M}_{\text{He}})$ and $z_{\text{He}}=\mathcal{M}_{\text{Ne}}R/(\mathcal{M}_{\text{Ne}}+\mathcal{M}_{\text{He}})$, with $\mathcal{M}_{\text{Ne}}$ and $\mathcal{M}_{\text{He}}$ the nuclear masses, and the dipole operator along the polarization reads
\begin{equation}\label{eq:dipole_operator}
  \hat{\mu}_\epsilon(R)=\mathcal{Q}_{\text{red}}R-|e|\hat{z},
\end{equation}

\noindent where
\begin{equation*}
  \mathcal{Q}_{\text{red}}=\frac{\mathcal{Z}_{\text{He}}\mathcal{M}_{\text{Ne}}-\mathcal{Z}_{\text{Ne}}\mathcal{M}_{\text{He}}}{\mathcal{M}_{\text{Ne}}+\mathcal{M}_{\text{He}}}\,|e|=0.9964\,|e|
\end{equation*}

\noindent is the reduced nuclear charge.

\subsection{Multistate reduction}
\label{sec:pecs}

The adiabatic electronic states $|\xi_i(R)\rangle$, which depend parametrically on $R$, with wave functions $\xi_i(z;R)=\langle z|\xi_i(R)\rangle$ and the PECs $\epsilon_i(R)$ solve
\begin{equation}\label{eq:TISE_BO}
  \hat{H}_{\text{el}}(z;R)\,\xi_i(z;R)=\epsilon_i(R)\,\xi_i(z;R),\qquad i=1,\dots,N_{\text{PECs}} .
\end{equation}
The ten lowest states are retained ($N_{\text{PECs}}=10$); states $\xi_5,\dots,\xi_{10}$ in most of the $R$ range represent a discretized continuum. The PECs are shown in Fig.~2 of Ref.~\cite{Mendez2025}. Once the electronic coordinate has been integrated out, the electronic factor of the Hilbert space is $\mathbb{C}^{N_{\text{PECs}}}$, spanned by the kets $|\xi_i(R)\rangle$, and every electronic operator is an $N_{\text{PECs}}\times N_{\text{PECs}}$ matrix that depends parametrically on $R$. The label $R$ is omitted when no confusion arises. In this basis the dipole operator is
\begin{eqnarray}\label{eq:dipole_matrix}
  \hat{\mu}_{\text{BOA}}(R) &=& \sum_{i,j=1}^{N_{\text{PECs}}}\mu^{\text{BOA}}_{ij}(R)\,|\xi_i(R)\rangle\langle\xi_j(R)|, \nonumber\\
  \mu^{\text{BOA}}_{ij}(R) &=& \langle\xi_i(R)|\hat{\mu}_\epsilon(R)|\xi_j(R)\rangle,
\end{eqnarray}
with PDMs $\mu^{\text{BOA}}_{ii}$ and transition dipole moments (TDMs) $\mu^{\text{BOA}}_{ij}$, $i\neq j$ (the superscript BOA refers to the Born--Oppenheimer approximation for the bare molecule, in which the states $\xi_i(R)$ are defined). The matrix $\mathbb{M}(R)\equiv[\mu^{\text{BOA}}_{ij}(R)]$ is real, symmetric and continuous in $R$. Neglecting the electronic nonadiabatic couplings $d_{ij}=\langle\xi_i|\partial_R\xi_j\rangle$, $i\neq j$ (Born--Oppenheimer approximation for the bare molecule), Eq.~\eqref{eq:PF_ham} becomes
\begin{eqnarray}\label{eq:PF_MLS_ham}
  \hat{H} &=& \hat{T}_R+\sum_{i=1}^{N_{\text{PECs}}}\epsilon_i(R)\,|\xi_i\rangle\langle\xi_i| + \frac{\hat{P}^2}{2}+\frac{\omega_0^2}{2}\hat{Q}^2 \nonumber\\
  &&+\, g\sqrt{\frac{2\omega_0}{\hbar}}\,\hat{Q}\,\hat{\mu}_{\text{BOA}}(R)+\hat{H}_{\text{DSE}}(R),
\end{eqnarray}
which defines the multilevel-system (MLS) model. Removing the PDMs means setting $\mu^{\text{BOA}}_{ii}\equiv0$, and removing the DSE means dropping the last term. The cavity-induced couplings between electronic states, mediated by $\hat{Q}$, are retained exactly in Eq.~\eqref{eq:PF_MLS_ham}; only the bare electronic couplings $d_{ij}$ are neglected.

\subsection{Truncated and full dipole self-energy}
\label{sec:dse}

In a truncated electronic space, the DSE term admits two inequivalent representations. If the square of the dipole operator is evaluated before the projection, the \emph{full dipole self-energy} (fDSE) is obtained,
\begin{eqnarray}\label{eq:fDSE}
  \hat{H}_{\text{fDSE}} &=& \frac{g^2}{\hbar\omega_0}\hat{\mathcal{D}}(R), \nonumber\\
  \big(\mathbb{D}\big)_{ij} &=& \langle\xi_i(R)|\hat{\mu}_\epsilon^{\,2}|\xi_j(R)\rangle .
\end{eqnarray}
If, instead, the dipole operator is projected first and then squared, the \emph{truncated dipole self-energy} (tDSE) follows,
\begin{eqnarray}\label{eq:tDSE}
  \hat{H}_{\text{tDSE}} &=& \frac{g^2}{\hbar\omega_0}\hat{\mu}_{\text{BOA}}^{\,2}(R), \nonumber\\
  \big(\mathbb{M}^2\big)_{ij} &=& \sum_{k=1}^{N_{\text{PECs}}}\mu^{\text{BOA}}_{ik}\mu^{\text{BOA}}_{kj} .
\end{eqnarray}
The tDSE preserves the completed-square structure of Eq.~\eqref{eq:completed_square} with the same projected dipole in the linear and quadratic terms; it is the form obtained by applying the PZW transformation to the truncated minimal-coupling Hamiltonian~\cite{DiStefano2019,Taylor2020,Taylor2024} and the one used in parametrized QED (pQED)~\cite{Weight2023}. The fDSE corresponds to projecting the full length-gauge Hamiltonian onto the retained electronic space, $\hat{H}\to\hat{\mathcal{P}}\hat{H}\hat{\mathcal{P}}$ with $\hat{\mathcal{P}}=\sum_{i=1}^{N_{\text{PECs}}}|\xi_i\rangle\langle\xi_i|$, so that $\hat{\mu}_\epsilon^{\,2}$ is projected as a whole. Inserting the resolution of the identity between the two dipole factors, the two forms are related by
\begin{eqnarray}\label{eq:mu2_closure}
  \mathbb{D} &=& \mathbb{M}^2+e^2\Delta\mathbb{M}^2, \nonumber\\
  (\Delta\mathbb{M}^2)_{ij} &=& \sum_{k>N_{\text{PECs}}}z_{ik}z_{kj}, \nonumber\\
  z_{ij} &=& \langle\xi_i|\hat{z}|\xi_j\rangle,
\end{eqnarray}
where only the electronic part of Eq.~\eqref{eq:dipole_operator} contributes, because the nuclear part is proportional to the identity. Since $i,j\le N_{\text{PECs}}<k$, only transition elements $z_{ik}$, $i\neq k$, between retained and excluded states enter $\Delta\mathbb{M}^2$, and never the permanent dipoles $z_{ii}$, which are the only elements changed by a shift of the origin. The matrix $\Delta\mathbb{M}^2$ is therefore independent of the dipole origin; it is also positive semidefinite and vanishes as $N_{\text{PECs}}\to\infty$. Its diagonal is bounded by the missing oscillator strength of the TRK sum rule (Appendix~\ref{app:trk}),
\begin{eqnarray}\label{eq:Delta_trk_bound}
  (\Delta\mathbb{M}^2)_{ii} &\le& \frac{\hbar^2}{2\mathcal{M}_e}\,\frac{1-\mathcal{F}_i^{(N_{\text{PECs}})}}{\epsilon_{N_{\text{PECs}}+1}-\epsilon_i}, \nonumber\\
  \mathcal{F}_i^{(N_{\text{PECs}})} &=& \sum_{j\le N_{\text{PECs}}}f_{ij},
\end{eqnarray}
with $f_{ij}$ the oscillator strengths and $\mathcal{M}_e$ the electron mass. Because the partial sums obey $\sum_{i\le N_{\text{PECs}}}\mathcal{F}_i^{(N_{\text{PECs}})}=0$ (the trace of a commutator of finite matrices vanishes), the missing oscillator strength accumulates in the last retained state, and the difference between fDSE and tDSE is concentrated in the row and column of $\xi_{N_{\text{PECs}}}$ (Sec.~\ref{sec:res_dipoles}).

Both forms share the property that guarantees gauge consistency. For a charged molecule the dipole origin is arbitrary~\cite{Haugland2020}: a constant shift changes $\mathbb{M}\to\mathbb{M}+\delta\mu\,\mathbb{I}$ with $\delta\mu\neq0$. As for the untruncated Hamiltonian~Eq.~\eqref{eq:PF_ham}~\cite{Rokaj2018,schafer_relevance_2020,DiStefano2019,Taylor2020}, the spectrum of the truncated Hamiltonian is invariant under this shift if and only if the DSE coefficient equals $\eta_{\text{req}}\equiv g^2/\hbar\omega_0$, and this holds for both truncated forms of the DSE (Sec.~S1 of the Supplemental Material~\cite{SM}). The requirement already holds at fixed $R$, where any other coefficient makes the spectrum depend on the arbitrary origin. With explicit nuclear dynamics its violation has a further consequence: the PDMs depend on $R$, and for a charged molecule they include the origin-dependent nuclear term $\mathcal{Q}_{\text{red}}R$, so that the residual becomes a potential along the dissociation coordinate that either removes the ground state or adds a spurious confinement. The DSE therefore becomes indispensable as soon as the nuclei move.

\subsection{Two-level and Rabi limits}
\label{sec:Rabi_model}

For $N_{\text{PECs}}=2$ (two-level system (TLS)), with the Pauli matrices in the basis $\{|\xi_1\rangle,|\xi_2\rangle\}$, $\hat{\sigma}_1=|\xi_1\rangle\langle\xi_2|+|\xi_2\rangle\langle\xi_1|$ and $\hat{\sigma}_3=|\xi_2\rangle\langle\xi_2|-|\xi_1\rangle\langle\xi_1|$, the identity $\hat{\sigma}_0=|\xi_1\rangle\langle\xi_1|+|\xi_2\rangle\langle\xi_2|$, and the symmetric and antisymmetric PDMs $\mu^{\text{BOA}}_{\pm}=(\mu^{\text{BOA}}_{22}\pm\mu^{\text{BOA}}_{11})/2$, the dipole matrix is $\mathbb{M}=\mu^{\text{BOA}}_{+}\hat{\sigma}_0+\mu^{\text{BOA}}_{-}\hat{\sigma}_3+\mu^{\text{BOA}}_{12}\hat{\sigma}_1$ and the tDSE becomes
\begin{equation}\label{eq:tls_M2}
  \begin{split}
  \mathbb{M}^2={}&\big[(\mu^{\text{BOA}}_{+})^2+(\mu^{\text{BOA}}_{-})^2+(\mu^{\text{BOA}}_{12})^2\big]\hat{\sigma}_0\\
  &+2\mu^{\text{BOA}}_{+}\mu^{\text{BOA}}_{-}\hat{\sigma}_3+2\mu^{\text{BOA}}_{+}\mu^{\text{BOA}}_{12}\hat{\sigma}_1 .
  \end{split}
\end{equation}
In atomic models without PDMs, $\mathbb{M}^2=(\mu^{\text{BOA}}_{12})^2\hat{\sigma}_0$ and the DSE is a constant energy offset that is usually discarded~\cite{ruggenthaler_understanding_2023}. Two features of the molecular case invalidate this argument. With nuclear motion, $(g^2/\hbar\omega_0)[\mu^{\text{BOA}}_{12}(R)]^2$ is a potential that modifies the nuclear dynamics; and with PDMs the DSE acquires the $\hat{\sigma}_3$ and $\hat{\sigma}_1$ components of Eq.~\eqref{eq:tls_M2}, which shift the transition energy and add a coupling proportional to $\mu^{\text{BOA}}_{+}\mu^{\text{BOA}}_{12}$. The latter depends on the dipole origin, and only the full combination of Eq.~\eqref{eq:PF_MLS_ham} with the DSE coefficient $\eta_{\text{req}}$ is origin independent.

The quantum Rabi model is recovered by setting $\mu^{\text{BOA}}_{\pm}=0$, discarding the DSE, and fixing $R$~\cite{rabi1937space,braak2011integrability},
\begin{equation}\label{eq:PF_Rabi_h}
  \hat{H}_{\text{Rabi}}=\frac{\Delta\epsilon_{12}}{2}\hat{\sigma}_3+\hbar\omega_0\hat{a}^{\dagger}\hat{a}+g\mu^{\text{BOA}}_{12}\,(\hat{a}+\hat{a}^{\dagger})\,\hat{\sigma}_1,
\end{equation}
with $\Delta\epsilon_{12}=\epsilon_2-\epsilon_1$; the Jaynes--Cummings model follows from the rotating-wave approximation, valid for $g|\mu^{\text{BOA}}_{12}|\ll\hbar\omega_0$ near resonance~\cite{jaynes1963comparison}. The molecular Hamiltonian departs from these models in three respects that are essential for the dynamics: $\Delta\epsilon_{12}(R)$ and $\mu^{\text{BOA}}_{ij}(R)$ depend on a dynamical nuclear coordinate; the PDMs grow linearly with $R$; and further electronic states, with their DSE, are retained. A symmetry consideration is also relevant. The Rabi Hamiltonian of Eq.~\eqref{eq:PF_Rabi_h} conserves the parity of the total number of excitations~\cite{braak2011integrability},
\begin{equation}\label{eq:parity}
  \hat{\Pi}=\exp\!\Big[\mathrm{i}\pi\Big(\hat{a}^{\dagger}\hat{a}+\frac{\hat{\sigma}_0+\hat{\sigma}_3}{2}\Big)\Big]=-\hat{\Pi}_{\text{ph}}\hat{\sigma}_3,
\end{equation}
where $\hat{\Pi}_{\text{ph}}=\exp(\mathrm{i}\pi\hat{a}^{\dagger}\hat{a})$ is the photon parity, which maps $\hat{a}\to-\hat{a}$, and $(\hat{\sigma}_0+\hat{\sigma}_3)/2=|\xi_2\rangle\langle\xi_2|$ counts the electronic excitation. Since $\hat{\Pi}$ changes the sign of both $\hat{a}+\hat{a}^{\dagger}$ and $\hat{\sigma}_1$, the interaction term is invariant, and for an initial state with a definite parity, such as $|0\rangle\otimes|\xi_2\rangle$, $\langle\hat{a}\rangle(t)\equiv0$ at all times, also with nuclear motion (with the $d_{ij}$ neglected, since the nonadiabatic term $d_{12}\partial_R$ changes sign under $\hat{\Pi}$) and with absorbing boundaries that act on $R$ only. In the presence of PDMs, the terms $\mu^{\text{BOA}}_{\pm}(R)(\hat{a}+\hat{a}^{\dagger})$ break this symmetry: a nonvanishing $\langle\hat{a}\rangle$ is then allowed and, as shown in Sec.~\ref{sec:res_coherence}, it is dominated by the static polarization field.

\subsection{Hierarchy of models}
\label{sec:hierarchy}

\begin{figure}[!tbp]
\centering
\includegraphics{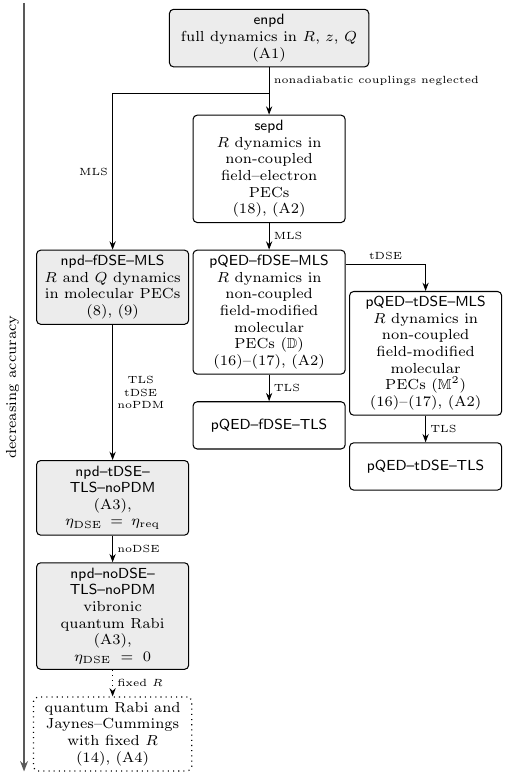}
\caption{\label{fig:models}Hierarchy of the models used in this work, labeled as \textsf{treatment--DSE--space(--noPDM)} (Sec.~\ref{sec:hierarchy}). The depth of each box indicates the accuracy of the model: the lower the box, the more approximations the model contains and the less accurate it is. Each arrow adds the approximation written next to it (nonadiabatic couplings neglected, common to all the models below \textsf{enpd}: $d_{ij}=0$ in \textsf{npd}, couplings between polaritonic surfaces in \textsf{sepd} and \textsf{pQED}; \textsf{MLS}: $N_{\text{PECs}}=10$; fixed $R$; \textsf{TLS}: $N_{\text{PECs}}=2$; \textsf{tDSE} instead of \textsf{fDSE}; \textsf{noDSE}; \textsf{noPDM}: $\mu^{\text{BOA}}_{ii}\equiv0$). Each box indicates the coordinates that the model propagates and the equations that define its Hamiltonian (Appendix~\ref{app:hamiltonians}). Shaded boxes are the wave-packet propagations performed here with MCTDH; the models in unshaded boxes reduce the problem to a one-dimensional dynamics in $R$ on polaritonic PECs, but are used here only for the PECs (and, for \textsf{pQED}, the branch weights of the initial state). The dotted box contains the standard models of quantum optics, not used in this work, which follow from \textsf{npd--noDSE--TLS--noPDM} at fixed $R$ (the Jaynes--Cummings model after the rotating-wave approximation).}
\end{figure}

Figure~\ref{fig:models} maps the models used in this work and the approximations that connect them. The approximate models are labeled \textsf{treatment--DSE--space(--noPDM)}: the \emph{treatment} keyword (\textsf{enpd}, \textsf{sepd}, \textsf{npd}, or \textsf{pQED}) indicates which degrees of freedom are propagated, the \emph{DSE} keyword (\textsf{fDSE}, \textsf{tDSE}, or \textsf{noDSE}) how the dipole self-energy is handled, the \emph{space} keyword (\textsf{MLS} or \textsf{TLS}) the electronic space, and the optional suffix \textsf{noPDM} that the permanent dipoles are discarded. Each keyword encodes one approximation with respect to the reference full model \textsf{enpd}.

The treatment field indicates which degrees of freedom are propagated: in \textsf{pQED} the electron--photon problem is diagonalized at fixed $R$ in the photon-number basis (Eqs.~\eqref{eq:An} and \eqref{eq:Cn}), so that only the nuclear coordinate remains dynamical and the problem reduces to a one-dimensional time-dependent Schr\"odinger equation in $R$ on the polaritonic PECs (Eq.~\eqref{eq:H_pec}); in \textsf{npd} (nucleus--photon dynamics) the electronic coordinate is expanded in the adiabatic basis and the nuclear and photonic coordinates are propagated together (Eq.~\eqref{eq:PF_MLS_ham}). The DSE field is \textsf{fDSE} (Eq.~\eqref{eq:fDSE}), \textsf{tDSE} (Eq.~\eqref{eq:tDSE}), or \textsf{noDSE} (DSE neglected), and the space keyword is \textsf{MLS} ($N_{\text{PECs}}=10$) or \textsf{TLS} ($N_{\text{PECs}}=2$). The optional suffix \textsf{noPDM} indicates that the PDMs are set to zero, $\mu^{\text{BOA}}_{ii}\equiv0$; its absence implies that they are retained. With this suffix, the TLS models become the vibronic (vibrationally coupled) Rabi model, i.e., Eq.~\eqref{eq:PF_Rabi_h} with nuclear motion: \textsf{npd--noDSE--TLS--noPDM} is this model without the DSE (DSE coefficient $\eta_{\text{DSE}}=0$ in Eq.~\eqref{eq:H_vRabi}), and \textsf{npd--tDSE--TLS--noPDM} adds the term $(g^2/\hbar\omega_0)[\mu^{\text{BOA}}_{12}(R)]^2\hat{\sigma}_0$ of Eq.~\eqref{eq:tls_M2} ($\eta_{\text{DSE}}=\eta_{\text{req}}$). Both conserve the parity of Eq.~\eqref{eq:parity}, and their dynamics is discussed in Sec.~S4 of the Supplemental Material~\cite{SM}.

Two models, \textsf{enpd} and \textsf{sepd}, retain the complete electronic space, so that neither the DSE nor the space field applies. The model \textsf{enpd} (electron--nucleus--photon dynamics) propagates Eq.~\eqref{eq:PF_ham} directly in the three coordinates $(R,z,Q)$, without further approximations, and serves as the reference for the dynamics (Sec.~\ref{sec:methods_full}). The full model includes all nonadiabatic couplings, which are neglected in every other model. The model \textsf{sepd} (simultaneous electron--photon diagonalization) diagonalizes the electron--photon Hamiltonian at fixed $R$ (Eq.~\eqref{eq:V2D}) on a two-dimensional grid in $(z,Q)$, without electronic truncation, so that the DSE is evaluated with the exact operator $\hat{z}^2$; as in \textsf{pQED}, only $R$ then remains to be propagated (Eq.~\eqref{eq:H_pec}). In this work, \textsf{sepd} and \textsf{pQED} are used for the polaritonic PECs, and \textsf{pQED} also for the branch weights of the initial state (Sec.~\ref{sec:methods_pecs}); neither is used for nuclear propagations.

%% file: section_methods.tex
\section{Methods}
\label{sec:methods}

\subsection{Electronic structure and molecular PECs}
\label{sec:methods_tise}

Equation~\eqref{eq:TISE_BO} was solved with the finite-element package FEMTISE~\cite{mendez2024femtise,bezanson2017julia}, which casts the weak form of the Sturm--Liouville problem into a generalized eigenvalue problem~\cite{sun2016finite} and yields variational upper bounds to the exact eigenvalues. Second-order Lagrange elements with $\Delta z=0.02$~a.u.\ on $z\in[-15,15]$~a.u.\ were used, on a grid of 1981 internuclear distances with $\Delta R=0.01$~a.u.\ in $[0.2,20]$~a.u. The matrices $z_{ij}$, $(z^2)_{ij}$, $\mu^{\text{BOA}}_{ij}$, and $(\mathbb{D})_{ij}$ were computed with the same Gauss quadrature and brought to the center-of-mass origin of Eq.~\eqref{eq:dipole_operator} analytically. All parameters are collected in Table~SII of the Supplemental Material~\cite{SM}. The molecular PECs $\epsilon_i(R)$ and the corresponding electronic states $|\xi_i(R)\rangle$ were computed for $i=1,\dots,10$, which is sufficient to converge the polaritonic PECs and the dynamics for the couplings considered. For $g=0.01$~a.u., the lowest polaritonic PECs are indistinguishable from the molecular PECs shifted by $n\hbar\omega_0$ on the scale of the electronic gaps (Sec.~\ref{sec:res_pecs}).

\subsection{Polaritonic potential-energy curves}
\label{sec:methods_pecs}

The polaritonic-surface partition of the generalized Born--Huang expansion~\cite{BornHuang1954,ruggenthaler_understanding_2023} groups the photon with the electron and solves the conditional electron--photon problem at each $R$~\cite{Galego2015}. Then, the polaritonic PECs $E_\alpha(R)$ are the eigenvalues of Eq.~\eqref{eq:PF_MLS_ham} without the kinetic energy term $\hat{T}_R$. Two treatments were used to compute them.

\paragraph{Photon-number basis (\textsf{pQED}).} In the product basis $|n,\xi_i\rangle\equiv|n\rangle\otimes|\xi_i(R)\rangle$, $n=0,\dots,n_{\text{max}}$, the Hamiltonian is block tridiagonal in $n$~\cite{Weight2023},
\begin{eqnarray}
  \langle n,\xi_i|\hat{H}|n,\xi_k\rangle &=& \Big[\epsilon_i+\hbar\omega_0\big(n+\tfrac12\big)\Big]\delta_{ik} \nonumber\\
  && +\,\frac{g^2}{\hbar\omega_0}\mathbb{S}_{ik}, \label{eq:An}\\
  \langle n+1,\xi_i|\hat{H}|n,\xi_k\rangle &=& g\sqrt{n+1}\,\mu^{\text{BOA}}_{ik}, \label{eq:Cn}
\end{eqnarray}
with $\mathbb{S}=\mathbb{D}$ (fDSE) or $\mathbb{M}^2$ (tDSE), and all blocks with $|n_1-n_2|\ge2$ vanish. Its diagonalization gives $E_\alpha(R)$ and the eigenvectors $|\Phi_\alpha(R)\rangle=\sum_{i,n}c^\alpha_{in}(R)|n,\xi_i\rangle$. In the calculations, $N_{\text{PECs}}=10$ and $n_{\text{max}}=50$ were used. Because the field is displaced by $\simeq g\mu^{\text{BOA}}_{ii}/\hbar\omega_0$ and the PDMs grow with $R$, the number of photon states required increases with $R$. For $g=0.12$~a.u., $n_{\text{max}}=50$ converges the lowest branches to machine precision and the 40 lowest to $\sim10^{-6}$~a.u.\ for $R\le10$~a.u., the region explored by the packet before it reaches the absorbing potential; beyond it, up to $n_{\text{max}}=80$ is needed (Sec.~S2 of the Supplemental Material~\cite{SM}).

\paragraph{Simultaneous electron--photon diagonalization (\textsf{sepd}).} The conditional Hamiltonian was also solved directly with FEMTISE~\cite{mendez2024femtise} on a two-dimensional finite-element grid in $(z,Q)$, i.e., the eigenvalue problem at fixed $R$ of
\begin{equation}\label{eq:V2D}
  \begin{split}
  \hat{H}(z,Q;R)={}&\hat{H}_{\text{el}}(z;R)+\frac{\hat{P}^2}{2}+\frac{\omega_0^2\hat{Q}^2}{2}+g\sqrt{\frac{2\omega_0}{\hbar}}\,\hat{Q}\,\hat{\mu}_\epsilon(R)\\
  &+\frac{g^2}{\hbar\omega_0}\hat{\mu}_\epsilon^{\,2}(R),
  \end{split}
\end{equation}
with $\hat{H}_{\text{el}}$ the electronic Hamiltonian of Eq.~\eqref{eq:TISE_BO} and $\hat{\mu}_\epsilon(R)$ the dipole operator of Eq.~\eqref{eq:dipole_operator}, both acting directly on $z$. This representation has no electronic or photon-number truncation, and the DSE is evaluated with the exact $\hat{z}^2$; its accuracy is controlled by the grid ($500\times500$ nodes, $\Delta z=\Delta Q=0.1$~a.u., $z,Q\in[-25,25]$~a.u.) and was computed on 100 values of $R$ with $\Delta R=0.2$~a.u. The pQED and sepd calculations are two representations of the same conditional problem, which differ in whether the electronic space is reduced before (pQED) or after (sepd) the coupling to the cavity. They must therefore agree in the limit of complete electronic and photon-number bases, and their comparison at $N_{\text{PECs}}=10$ quantifies the effect of the electronic truncation (Sec.~\ref{sec:res_pecs}).

\subsection{Initial state and interaction quench}
\label{sec:initial_state}

The molecule is assumed to be in the vibrational ground state $\chi_1(R)$ of $\epsilon_1(R)$, to be promoted vertically to $\xi_2$, and to find the cavity mode in the length-gauge vacuum,
\begin{equation}\label{eq:initial_state}
  |\Psi(0)\rangle=\int\text{d}R\,\chi_1(R)\,|R\rangle\otimes|0\rangle\otimes|\xi_2(R)\rangle .
\end{equation}
The packet has $\langle R\rangle=1.679$~a.u.\ and $\sigma_R=0.118$~a.u., and its energy on $\epsilon_2$ lies 0.39~a.u.\ above the dissociation limit of $\epsilon_2$, so that the bare dynamics is dissociative with an asymptotic velocity $v\simeq0.011$~a.u. The cavity frequency, $\hbar\omega_0=0.645$~a.u.\ ($17.6$~eV), equals the vertical excitation energy at the equilibrium distance, $\epsilon_2-\epsilon_1=\hbar\omega_0$ at $R_{\text{eq}}=1.6654$~a.u. The state $|0,\xi_2\rangle$ is therefore degenerate with $|1,\xi_1\rangle$, i.e., the dynamics starts on the resonant avoided crossing of the polaritonic PECs (see Fig.~\ref{fig:pecs}(a)).

Since Eq.~\eqref{eq:initial_state} is not an eigenstate of $\hat{H}$, energy is deposited in the system not only by the electronic excitation but also by the sudden coupling to the cavity. Because $\langle0|\hat{a}+\hat{a}^{\dagger}|0\rangle=0$, the energy of this interaction quench is
\begin{equation}\label{eq:quench_energy}
  \begin{split}
  E_{\text{q}}&=\langle\Psi(0)|\hat{H}_{\text{fDSE}}|\Psi(0)\rangle\\
  &=\frac{g^2}{\hbar\omega_0}\big\langle(\mathbb{D})_{22}\big\rangle_{\chi_1}=\hbar\omega_0\langle\hat{n}_{\text{PZW}}\rangle_0,
  \end{split}
\end{equation}
i.e., it equals the energy of the free photons contained in the length-gauge vacuum (Eq.~\eqref{eq:physical_photon}). With $\langle(\mathbb{D})_{22}\rangle_{\chi_1}\simeq3.63$~a.u., $E_{\text{q}}=5.6\times10^{-4}$~a.u.\ ($0.015$~eV) for $g=0.01$~a.u.\ and $7.9\times10^{-2}$~a.u.\ ($2.2$~eV) for $g=0.12$~a.u., that is, from $0.09\%$ to $12\%$ of the electronic excitation energy $\hbar\omega_0$ (Table~\ref{tab:couplings}).

Using the \textsf{pQED} treatment implies the projection the initial state of Eq.~\eqref{eq:initial_state} onto the polaritonic basis of \textsf{pQED--fDSE--MLS} and defines the branch weights $p_\alpha^{\chi_1}=\int\text{d}R\,|\chi_1(R)|^2|c^\alpha_{20}(R)|^2$, where $c^\alpha_{20}(R)=\langle0,\xi_2(R)|\Phi_\alpha(R)\rangle$ is the coefficient of $|0,\xi_2\rangle$ in the eigenvector $|\Phi_\alpha(R)\rangle$.

\subsection{Coupled nucleus--photon dynamics}
\label{sec:methods_dynamics}

The time-dependent Schr\"odinger equation for Eq.~\eqref{eq:PF_MLS_ham} with fDSE (model \textsf{npd--fDSE--MLS}) was solved with the multi-set MCTDH method as implemented in the Heidelberg package~\cite{Meyer1990,Beck2000,MeyerMCTDHpackage}. The wave function is expanded, for each electronic state, in products of time-dependent single-particle functions (SPFs) of $R$ and $Q$. The wave function was propagated up to 30~fs for eight couplings between $g=0.01$ and $0.12$~a.u. The dissociation fragments that reach large $R$ were absorbed by a complex absorbing potential (CAP), designed to reflect less than $10^{-11}$ of the probability in the energy range of the packet. The integrator tolerances, the convergence tests and other details are reported in Sec.~S3 of the Supplemental Material~\cite{SM}.

\subsection{Full electron--nucleus--photon dynamics}
\label{sec:methods_full}

To assess the electronic truncation, the PF Hamiltonian of Eq.~\eqref{eq:PF_ham} was also propagated for $g=0.12$~a.u. (the strongest coupling value)\ with the electron treated in full as a third MCTDH mode (model \textsf{enpd}), with a single-set wave function $\Psi(R,z,Q,t)$. The electronic coordinate is measured from the internuclear midpoint, as in the finite-element calculations. This choice keeps the electron--nucleus potential identical to the one that defines the states $\xi_i$, which allows a direct validation against the finite-element results, and places the nuclei symmetrically at $z=\mp R/2$, so that a symmetric electronic box with identical absorbing potentials at both ends contains both nuclei over the whole range of $R$ (with origin at the nuclear center of mass, the He nucleus would reach $z\simeq17$~a.u.\ at $R=20$~a.u.). At fixed center of mass the mass-polarization terms $(\hbar^2s/\mathcal{M}_{\text{red}})\,\partial_R\partial_z-(\hbar^2s^2/2\mathcal{M}_{\text{red}})\,\partial_z^2$, with $s=(\mathcal{M}_{\text{Ne}}-\mathcal{M}_{\text{He}})/[2(\mathcal{M}_{\text{Ne}}+\mathcal{M}_{\text{He}})]=0.3345$, were included. These terms are absent in \textsf{npd--fDSE--MLS}, whose adiabatic states are computed at fixed $R$ with the same midpoint coordinate, so that the comparison between both models also contains the effect of the mass polarization, in addition to those of the continuum and of the non-adiabatic couplings $d_{ij}$. Since the nuclear center of mass lies at $z=-sR$ with respect to the midpoint, the electronic coordinate measured from the center of mass is $z+sR$, and the dipole operator of Eq.~\eqref{eq:dipole_operator}, still referred to the nuclear center of mass, reads $\hat{\mu}_\epsilon=\mathcal{Q}_{\text{mid}} R-|e|\hat{z}$ with $\mathcal{Q}_{\text{mid}}=\mathcal{Q}_{\text{red}}-s|e|=0.6619\,|e|$. The electron--nucleus potential was represented through its exact matrix elements in a sine DVR, the electrons emitted into the continuum were absorbed by a cubic potential in $z$, and the Franck--Condon state was written as a sum of products; the details and the tests of this representation are given in Sec.~S3 of the Supplemental Material~\cite{SM}.

\subsection{Observables}
\label{sec:observables}
The wave function is written as $\Psi(R,Q,t)=\sum_i\psi_i(R,Q,t)|\xi_i\rangle$, and the population of the electronic state $\xi_i$ is $p_{\xi_i}(t)=\iint|\psi_i|^2\,\text{d}R\,\text{d}Q$. Their sum is the norm $\mathcal{N}(t)=\langle\Psi(t)|\Psi(t)\rangle=\sum_ip_{\xi_i}(t)$, which decreases only through the absorbing potential. The nuclear and photonic densities projected on each electronic state are $\rho_i(R,t)=\int|\psi_i|^2\,\text{d}Q$ and $\rho_i(Q,t)=\int|\psi_i|^2\,\text{d}R$; the latter is the probability density along the mode coordinate $Q$, which in the length gauge represents the displacement field (Sec.~\ref{sec:pf_hamiltonian}). Two photon-number distributions are computed: the populations of the eigenstates of the length-gauge number operator $\hat{n}$, $p_n^{\text{fock}}(t)=\|\langle n|\Psi(t)\rangle\|^2$, and those of the eigenstates of the free-photon number operator $\hat{n}_{\text{PZW}}$ of Eq.~\eqref{eq:physical_photon}, $p_n^{\text{phys}}(t)=\|\langle n|\hat{U}_{\text{PZW}}^\dagger\Psi(t)\rangle\|^2$, where $\langle n|$ acts on the photonic factor and the norm is taken over the nuclear and electronic factors (Appendix~\ref{app:photon}).

Photonic expectation values are conditioned on survival, $\langle\hat{O}\rangle(t)=\langle\Psi|\hat{O}|\Psi\rangle/\mathcal{N}(t)$, and the intracavity second-order coherence of the free photons is $\mathcal{G}^{(2)}_{\text{PZW}}(0)=\langle\hat{a}_{\text{PZW}}^{\dagger\,2}\hat{a}_{\text{PZW}}^{2}\rangle/\langle\hat{n}_{\text{PZW}}\rangle^2$~\cite{glauber1963quantum,glauber1963coherent}. Since $\hat{a}_{\text{PZW}}^{\dagger\,2}\hat{a}_{\text{PZW}}^{2}=\hat{n}_{\text{PZW}}(\hat{n}_{\text{PZW}}-1)$, it is fixed by the mean and the variance $\sigma^2_{\text{PZW}}=\langle\hat{n}_{\text{PZW}}^2\rangle-\langle\hat{n}_{\text{PZW}}\rangle^2$ of the photon-number distribution, $\mathcal{G}^{(2)}_{\text{PZW}}(0)=1+(\sigma^2_{\text{PZW}}-\langle\hat{n}_{\text{PZW}}\rangle)/\langle\hat{n}_{\text{PZW}}\rangle^2$, and distinguishes sub-Poissonian ($<1$, $\sigma^2_{\text{PZW}}<\langle\hat{n}_{\text{PZW}}\rangle$), Poissonian ($=1$), and super-Poissonian ($>1$) statistics~\cite{Mandel1986,MandelWolf1995}. Reference values follow from the variance of each state: $1-1/n$ for a Fock state $|n\rangle$ ($\sigma^2_{\text{PZW}}=0$), $1$ for a coherent state ($\sigma^2_{\text{PZW}}=\langle\hat{n}_{\text{PZW}}\rangle$), $2$ for a thermal state ($\sigma^2_{\text{PZW}}=\langle\hat{n}_{\text{PZW}}\rangle+\langle\hat{n}_{\text{PZW}}\rangle^2$), and $3+1/\langle\hat{n}_{\text{PZW}}\rangle$ for a squeezed vacuum ($\sigma^2_{\text{PZW}}=2\langle\hat{n}_{\text{PZW}}\rangle+2\langle\hat{n}_{\text{PZW}}\rangle^2$)~\cite{gerry_introductory_2004}. Equivalently, in terms of the free-photon number distribution, $\mathcal{G}^{(2)}_{\text{PZW}}(0)=\sum_nn(n-1)\,p_n^{\text{phys}}\big/\big(\sum_nn\,p_n^{\text{phys}}\big)^2$. It turns out the $\mathcal{G}^{(2)}_{\text{PZW}}(0)$ is very sensitive to the photon distribution $p_n^{\text{phys}}$ and it must be carefully treated as described in Appendix~\ref{app:photon}.

The von Neumann entropy characterizes the correlations between the three subsystems (nucleus, photon, and electron), but a single bipartite entropy does not capture all of them. The quantity used here is the entanglement between $R$ and $Q$ \emph{within} each electronic state, obtained from the natural populations $\nu_k^{(i)}$ of the conditional state $\psi_i/\sqrt{p_{\xi_i}}$,
\begin{equation}\label{eq:vn_entropy}
  S_i(t)=-\sum_k\frac{\nu_k^{(i)}}{p_{\xi_i}}\ln\frac{\nu_k^{(i)}}{p_{\xi_i}}.
\end{equation}

Since $\psi_i/\sqrt{p_{\xi_i}}$ is a pure state of $R$ and $Q$, the reduced density matrices of both modes have the same spectrum, and $S_i$ is a proper entanglement measure for it~\cite{nielsen_quantum_2011}. When $p_{\xi_i}\lesssim10^{-7}$, the conditional state is dominated by the numerical error of the propagation, and $S_i$ has no physical meaning; the small negative values that this error produces are set to zero. With the weights $w_i=p_{\xi_i}/\mathcal{N}$, which add up to one, the entanglement generated within the electronic states is summarized by the weighted entropy $\bar{S}=\sum_iw_iS_i$. Since the $\xi_i(z;R)$ are orthonormal for each $R$, the reduced density matrix of the photon in the multistate model is $\rho_Q=\sum_iw_i\,\rho_Q^{(i)}$, where $\rho_Q^{(i)}$ is that of the conditional state $\psi_i/\sqrt{p_{\xi_i}}$ and has entropy $S_i$. The concavity of the von Neumann entropy and the upper bound on the entropy of a convex combination of density matrices then give, for the entropy $S_Q$ of the photon with the rest of the system~\cite{nielsen_quantum_2011},
\begin{equation}\label{eq:entropy_bounds}
  \bar{S}\le S_Q\le\bar{S}+H(w),\qquad H(w)=-\sum_iw_i\ln w_i.
\end{equation}
The difference $S_Q-\bar{S}=\sum_iw_i\,D\big(\rho_Q^{(i)}\|\rho_Q\big)$, where $D(\rho\|\sigma)=\mathrm{Tr}\,\rho(\ln\rho-\ln\sigma)$ is the relative entropy, measures how different the photonic states of the electronic channels are: it vanishes when all $\rho_Q^{(i)}$ are equal, and reaches its maximum, $H(w)$, when they are orthogonal, i.e., when the photon state identifies the electronic state.

All entropies are computed in the length gauge, i.e., with respect to the partition between the displacement-field mode and the molecule. Since the PZW transformation is not a product of local unitaries, entanglement measures depend on this partition~\cite{Stokes2022}; in the presence of large PDMs, part of the length-gauge entropy reflects the state-dependent static polarization field rather than free photons. Since only the natural populations of each $\psi_i$ are computed here, only the two bounds of Eq.~\eqref{eq:entropy_bounds} are available for $S_Q$ in \textsf{npd--fDSE--MLS}.

All entropies oscillate with the rotation period of the free-photon coherence $\langle\hat{a}_{\text{PZW}}\rangle$ (Sec.~\ref{sec:res_entanglement}). To show their slow evolution, each curve $y(t_k)$ is replaced by a local average as described in Appendix~\ref{app:photon}.

\subsection{Coupling strengths}
\label{sec:couplings}

The coupling strengths reachable in nanophotonic setups, and the very description of their coupling, are under debate~\cite{Pradilla2025}. To place the values used here, Table~\ref{tab:couplings} lists, for the eight couplings, the mode volume $V=2\pi\omega_0/g^2$ (atomic units) of Eq.~\eqref{eq:g_def}, the normalized coupling at the resonance $\eta_{\text{R}}=g|\mu^{\text{BOA}}_{12}(R_{\text{eq}})|/\hbar\omega_0$, with $\mu^{\text{BOA}}_{12}(R_{\text{eq}})=0.875$~a.u., the first-order vacuum Rabi splitting $2g|\mu^{\text{BOA}}_{12}(R_{\text{eq}})|$, and the quench energy $E_{\text{q}}$ of Eq.~\eqref{eq:quench_energy}.

According to the usual criterion $\eta_{\text{R}}\gtrsim0.1$~\cite{forn2019ultrastrong,frisk2019ultrastrong}, the two largest couplings enter the perturbative USC regime, and all couplings are far below the deep-strong regime $\eta_{\text{R}}\gtrsim1$. The experimental relevance of these values is discussed in Sec.~\ref{sec:disc_feasibility}. For the two largest couplings, $E_{\text{q}}$ exceeds 1~eV and becomes comparable to the electronic excitation energies of the molecule, so that the cavity is no longer a small perturbation of the bare dissociation and higher electronic states can be reached (Sec.~\ref{sec:res_densities}).

\begin{table}[htbp]
\caption{\label{tab:couplings}Light--matter couplings $g$, mode volume $V$, normalized coupling at the resonance $\eta_{\text{R}}$, first-order vacuum Rabi splitting $2g|\mu^{\text{BOA}}_{12}(R_{\text{eq}})|$, and quench energy $E_{\text{q}}$ of Eq.~\eqref{eq:quench_energy}. The couplings are listed with the precision used in the calculations; elsewhere they are quoted rounded, $g=0.01$, $0.014$, $0.02$, $0.03$, $0.04$, $0.06$, $0.08$, and $0.12$~a.u.}
\begin{ruledtabular}
\sisetup{group-digits=none}
\begin{tabular}{S[table-format=1.6]S[table-format=1.3]S[table-format=1.3]S[table-format=1.2]S[table-format=1.3]}
{$g$~(a.u.)} & {$V$~(nm$^3$)} & {$\eta_{\text{R}}$} & {$2g|\mu^{\text{BOA}}_{12}|$~(eV)} & {$E_{\text{q}}$~(eV)} \\
\hline
0.010000 & 6.01  & 0.014 & 0.48 & 0.015 \\
0.014236 & 2.96  & 0.019 & 0.68 & 0.031 \\
0.020267 & 1.46  & 0.028 & 0.97 & 0.063 \\
0.028853 & 0.721 & 0.039 & 1.37 & 0.13 \\
0.041077 & 0.356 & 0.056 & 1.96 & 0.26 \\
0.058480 & 0.176 & 0.079 & 2.79 & 0.53 \\
0.083255 & 0.087 & 0.113 & 3.96 & 1.07 \\
0.118525 & 0.043 & 0.161 & 5.65 & 2.16 \\
\end{tabular}
\end{ruledtabular}
\end{table}

%% file: section_results_small.tex
\section{Results}
\label{sec:results}

The results from this section are presented in the order of the different quantitites one needs to perform the dynamics of the system in the different approaches. We begin with describing the matrix elements of the dipole operator, discussing the permanent and transition terms, and then construct the polaritonic PECs. The dynamics is then analyzed in terms of the nuclear and photonic densities, and the photon statistics. We also compare the results of the reduced models with the full electron--nucleus--photon dynamics. Finally, we discuss the bipartite entanglement dynamics in both the full approach and the reduced models.

\subsection{Dipole moments and dipole self-energy matrices}
\label{sec:res_dipoles}

\begin{figure*}[tbp]
  \centering
  \includegraphics{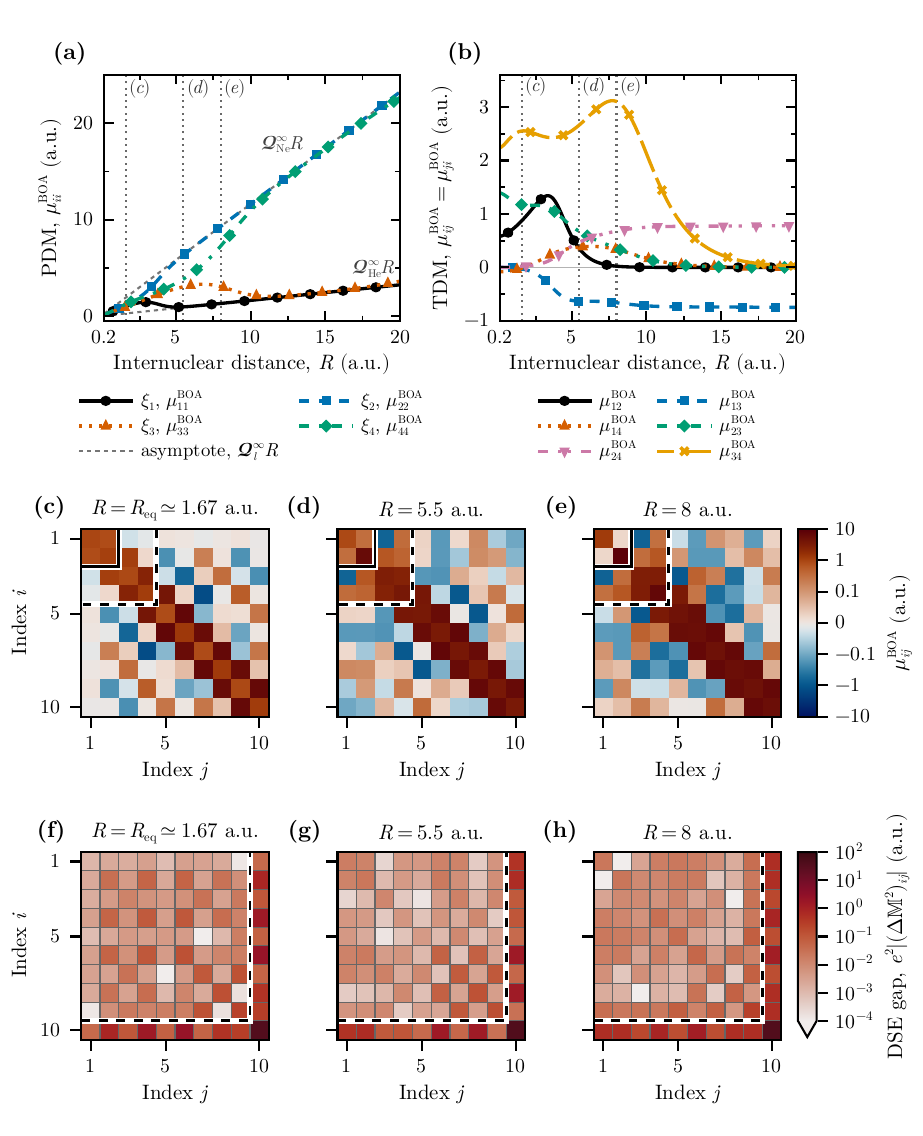}
  \caption{\label{fig:dipoles} Dipole matrix elements and DSE gap. Dipole moments referred to the nuclear center of mass, in the real electronic gauge continuous in $R$. (a) PDMs $\mu^{\text{BOA}}_{ii}$, $i\le4$ (a.u.), as functions of the internuclear distance $R$ (a.u.), with the linear asymptotes $\mathcal{Q}^{\infty}_{\text{He}}R$ and $\mathcal{Q}^{\infty}_{\text{Ne}}R$ of Eq.~\eqref{eq:pdm_asymptotes} (gray). (b) TDMs $\mu^{\text{BOA}}_{ij}$, $i<j\le4$ (a.u.); the matrix is real and symmetric. Vertical dotted lines mark the distances of panels (c)--(e). (c)--(e) Matrix $\mu^{\text{BOA}}_{ij}$ (a.u.) for $N_{\text{PECs}}=10$ at (c) $R=R_{\text{eq}}\simeq1.67$~a.u., (d) $R=5.5$~a.u., and (e) $R=8$~a.u., on a common diverging scale, linear for $|\mu^{\text{BOA}}_{ij}|<0.1$~a.u.\ and logarithmic outside (red, positive; blue, negative). The solid and dashed frames delimit the $N_{\text{PECs}}=2$ and $N_{\text{PECs}}=4$ blocks. (f)--(h) Difference between the two evaluations of the quadratic DSE term, $e^2|(\Delta\mathbb{M}^2)_{ij}|=|(\mathbb{D})_{ij}-(\mathbb{M}^2)_{ij}|$ (Eq.~\eqref{eq:mu2_closure}) (a.u.), at the same distances as (c)--(e). Row $i$ and column $j$ label the electronic states $\xi_i$ and $\xi_j$ ordered by energy. Logarithmic single-hue color scale common to the three panels (darker, larger difference); values below $10^{-4}$~a.u.\ are clipped. The dashed lines isolate the row and column of the last retained state, $i=N_{\text{PECs}}$, where the error accumulates.}
\end{figure*}

Figure~\ref{fig:dipoles} shows the dipole matrix elements of Eq.~\eqref{eq:dipole_matrix}, that enter into Eq.~\eqref{eq:PF_MLS_ham}.
At large distances $R$ between the nuclei the electron is mostly localized in each center $l\in\{\text{Ne},\text{He}\}$. Consequently, the electron position can be approximated as $\langle\hat{z}\rangle_i\to z_l$ and the PDMs approach the line $\mathcal{Q}^{\infty}_lR$, where the effective asymptotic charges are
\begin{eqnarray}\label{eq:pdm_asymptotes}
  \mathcal{Q}^{\infty}_{\text{Ne}} &=& \mathcal{Q}_{\text{red}}+|e|\frac{\mathcal{M}_{\text{He}}}{\mathcal{M}_{\text{Ne}}+\mathcal{M}_{\text{He}}}=1.1619\,|e|, \nonumber\\
  \mathcal{Q}^{\infty}_{\text{He}} &=& \mathcal{Q}_{\text{red}}-|e|\frac{\mathcal{M}_{\text{Ne}}}{\mathcal{M}_{\text{Ne}}+\mathcal{M}_{\text{He}}}=0.1619\,|e|.
\end{eqnarray}
States $\xi_1$ and $\xi_3$ follow $\mathcal{Q}^{\infty}_{\text{He}}R$ and states $\xi_2$ and $\xi_4$ follow $\mathcal{Q}^{\infty}_{\text{Ne}}R$ (Fig.~\ref{fig:dipoles}(a)), as is expected from the electron position $\langle\hat{z}\rangle_i$. Since $\mathcal{Q}^{\infty}_{\text{Ne}}-\mathcal{Q}^{\infty}_{\text{He}}=|e|$, the antisymmetric PDM $\mu^{\text{BOA}}_{-}= (\mu^{\text{BOA}}_{22}-\mu^{\text{BOA}}_{11})/2$ goes to $|e|R/2$, i.e., the dipole of one elementary charge transferred between the two centers.

The TDM $\mu^{\text{BOA}}_{12}$ reaches $1.34$~a.u.\ at $R\simeq3.4$~a.u., where both states share the charge, and vanishes at large $R$, where they are localized on different centers (Fig.~\ref{fig:dipoles}(b)). At $R_{\text{eq}}$, $\xi_2$ is strongly coupled to $\xi_3$ ($|\mu^{\text{BOA}}_{23}|=1.17$~a.u.) with a gap $\epsilon_3-\epsilon_2=0.60$~a.u.\ close to $\hbar\omega_0$, so that $|1,\xi_2\rangle$ and $|0,\xi_3\rangle$ are nearly resonant; Note that this important transition is not taken into account in the TLS (Sec.~\ref{sec:res_pecs}) and is considered the main error contribution in that approximation. On the other hand, the couplings of $\xi_1$ and $\xi_2$ with the pseudocontinuum states $\xi_5,\dots,\xi_{10}$ are less than $0.21$~a.u.

The full matrices (Figs.~\ref{fig:dipoles}(c)--\ref{fig:dipoles}(e)) show that the PDMs dominate at large $R$, as the growth of the diagonal elements shows. The dipole operator mixes all retained states, but at the equilibrium distance the TLS block couples mainly to $\xi_3$ (through $\mu^{\text{BOA}}_{23}$), which in turn couples to $\xi_4$ (through $\mu^{\text{BOA}}_{34}$), effectively producing a chain of transitions (Fig.~\ref{fig:dipoles}(c)). At $R=5.5$~a.u., the TLS block is strongly coupled to both $\xi_3$ and $\xi_4$ (Fig.~\ref{fig:dipoles}(d)), so that a four-state model would be needed at this distance. For $R=8$~a.u., the transition dipoles within the TLS block vanish, because $\xi_1$ and $\xi_2$ are localized on different centers (Fig.~\ref{fig:dipoles}(e)). At the largest couplings, however, the dynamics populates the higher electronic states (Sec.~\ref{sec:res_densities}), so that the TLS is no longer adequate.

We will now discuss the two expressions for the DSE. As one can see in Eq.~\eqref{eq:mu2_closure}, they differ by $e^2\Delta\mathbb{M}^2$, shown in Figs.~\ref{fig:dipoles}(f)--\ref{fig:dipoles}(h). The ground state nearly saturates the TRK sum rule within the retained space, $\mathcal{F}_1^{(10)}=0.991$, and $(\Delta\mathbb{M}^2)_{11}=1.3\times10^{-3}$~a.u.\ at $R_{\text{eq}}$; the low-lying states have $(\Delta\mathbb{M}^2)_{ii}\le0.26$~a.u.\ for $i\le9$, consistent with the bound of Eq.~\eqref{eq:Delta_trk_bound}. In contrast, $\mathcal{F}_{10}^{(10)}=-8.06$ and $(\Delta\mathbb{M}^2)_{10,10}=37.9$~a.u.: the missing oscillator strength accumulates in the edge state, as required by the trace rule, and the error concentrates in its row and column at all distances (Appendix~\ref{app:trk}). For the pair of the TLS, $e^2|(\Delta\mathbb{M}^2)_{12}|$ is of order $10^{-2}$~a.u.\ (Figs.~\ref{fig:dipoles}(f)--\ref{fig:dipoles}(h)). The energetic consequence on the polaritonic PECs is small for the branches that carry the wave packet: the fDSE--tDSE difference of the two lowest branches stays below $10^{-3}$~a.u.\ even for $g=0.12$~a.u., scales as $g^2$, and leaves the position and the minimum gap of the resonant avoided crossing practically unchanged (Table~\ref{tab:pec_errors}). The branches with appreciable weight on $\xi_{10}$ are shifted by up to $\delta E_{10}\simeq0.8$~a.u.\ (Table~\ref{tab:dse_gap}) and are unphysical in either form; they carry a negligible fraction of the initial state (Sec.~\ref{sec:res_pecs}).

\subsection{Polaritonic potential-energy curves}
\label{sec:res_pecs}

\begin{figure*}[tbp]
  \centering
  \includegraphics{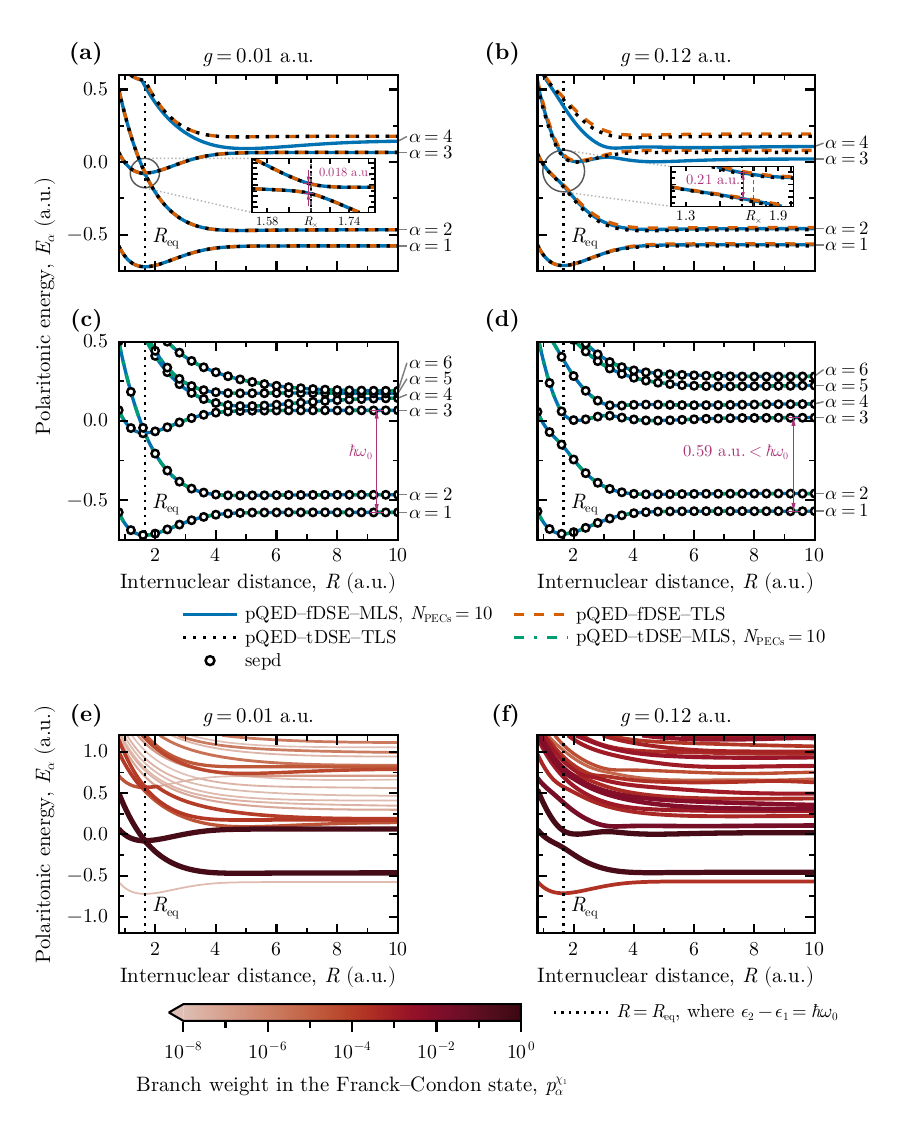}
  \caption{\label{fig:pecs}Polaritonic PECs $E_\alpha(R)$ (a.u.) as functions of the internuclear distance $R\in[0.8,10]$ (a.u.), for $g=0.01$~a.u.\ (left column) and $g=0.12$~a.u.\ (right column). (a,b) Multistate versus two-level models, branches $\alpha\le4$: \textsf{pQED--fDSE--MLS} with $N_{\text{PECs}}=10$ (solid), \textsf{pQED--fDSE--TLS} (dashed), and \textsf{pQED--tDSE--TLS} (dotted). The insets magnify the resonant avoided crossing between branches 2 and 3; the vertical bar gives the minimum gap, and $R_\times$ marks the equal-mixing point (Sec.~\ref{sec:res_pecs}). (c,d) pQED versus simultaneous electron--photon diagonalization, branches $\alpha\le6$: \textsf{pQED--fDSE--MLS} (solid), \textsf{pQED--tDSE--MLS} (dash-dotted), and \textsf{sepd} (open circles, at alternate nodes of the $\Delta R=0.2$~a.u.\ grid). Branches with $\alpha\ge7$ involve the discretized continuum, which the two representations do not sample with the same density of states (Sec.~S2 of the Supplemental Material~\cite{SM}). The vertical bars give $E_3-E_1$ at $R=9.3$~a.u., which equals $\hbar\omega_0$ for $g=0.01$~a.u.\ but is smaller for $g=0.12$~a.u., where branch 3 is mixed with $|0,\xi_3\rangle$ (Sec.~\ref{sec:res_pecs}). (e,f) Thirty lowest branches of \textsf{pQED--fDSE--MLS}, colored and thickened according to their weight $p_\alpha^{\chi_1}$ (Sec.~\ref{sec:initial_state}) in the Franck--Condon state (logarithmic color bar). In all panels, the dotted vertical line marks the equilibrium distance $R_{\text{eq}}$, where $\epsilon_2-\epsilon_1=\hbar\omega_0$.}
\end{figure*}

Figure~\ref{fig:pecs} summarizes the polaritonic PECs for the smallest and the largest coupling used in this work. In the polaritonic-surface partition, the vacuum Rabi splitting is directly visible~\cite{Galego2015}: branches 2 and 3 in the figure form the resonant avoided crossing between $|0,\xi_2\rangle$ and $|1,\xi_1\rangle$ at $R_{\text{eq}}$, with a gap that grows from $0.018$~a.u.\ for $g=0.01$~a.u.\ to $0.207$~a.u.\ for $g=0.12$~a.u. ($32\%$ of $\hbar\omega_0$)(insets of Figs.~\ref{fig:pecs}(a) and \ref{fig:pecs}(b)), close to the first-order estimate $2g|\mu^{\text{BOA}}_{12}(R_{\text{eq}})|$ of Table~\ref{tab:couplings}. The minimum of the gap stays at $R_{\text{eq}}$, within the grid spacing, for all couplings (Table~\ref{tab:pec_errors}).

The equal-mixing point $R_\times$ is defined as the position in which branches 2 and 3 carry equal weights of $|0,\xi_2\rangle$, $|c^{2}_{20}(R_\times)|^2=|c^{3}_{20}(R_\times)|^2$ (Sec.~\ref{sec:initial_state}). In a two-level crossing with an $R$-independent coupling, $R_\times$ coincides with the minimum of the gap; here the coupling $\mu^{\text{BOA}}_{12}$ depends on $R$ and more than two states contribute to the branches, so the minimum of the gap stays at $R_{\text{eq}}$ (Table~\ref{tab:pec_errors}) while $R_\times$ moves. Specifically it moves to larger $R$ as $g$ grows, because the cavity shifts the two diabatic states differently (mainly through the coupling of $\xi_2$ to $\xi_3$). $R_\times$ lies at $R=1.665$~a.u.\ ($\simeq R_{\text{eq}}$) for $g=0.01$~a.u.\ and at $1.738$~a.u.\ for $g=0.12$~a.u.\ (label $R_\times$ in the insets of Figs.~\ref{fig:pecs}(a) and \ref{fig:pecs}(b)); the displacement, $0.07$~a.u., remains smaller than the width $\sigma_R=0.118$~a.u.\ of the initial nuclear packet (Sec.~\ref{sec:initial_state}), so the packet is born on the crossing for all couplings.

Away from $R_{\text{eq}}$ the detuning $\Delta\epsilon_{12}-\hbar\omega_0$ between the electronic transition and the field grows rapidly ($\partial_R\Delta\epsilon_{12}=-0.45$~a.u.\ at $R_{\text{eq}}$). In the linear approximation, the detuning equals the vacuum Rabi gap at $|R-R_{\text{eq}}|\simeq0.04$~a.u.\ for $g=0.01$~a.u.\ and $\simeq0.46$~a.u.\ for $g=0.12$~a.u.; beyond this distance the $\xi_1$--$\xi_2$ transition is out of resonance with the cavity. In the TLS, where $\mu^{\text{BOA}}_{12}$ vanishes at large $R$, the branches then recover the bare PECs shifted by integer multiples of $\hbar\omega_0$ for all couplings (dashed and dotted lines in Figs.~\ref{fig:pecs}(a) and \ref{fig:pecs}(b)), up to a nearly rigid shift of $\approx0.015$~a.u.\ in fDSE at $g=0.12$~a.u. This does not hold in the MLS. There, the one-photon branch $|1,\xi_1\rangle$ remains mixed with $|0,\xi_3\rangle$ in the dissociation limit, because $\mu^{\text{BOA}}_{13}$ does not vanish ($0.72$~a.u.\ at $R=10$~a.u.) and, for $g=0.12$~a.u., $g|\mu^{\text{BOA}}_{13}|\simeq0.085$~a.u.\ is comparable to their detuning ($0.076$~a.u.). Branches 3 and 4 are thus shifted by $\approx0.04$~a.u.\ from the bare PECs (Fig.~\ref{fig:pecs}(b)), and at large $R$ the spacing $E_3-E_1$ equals $\hbar\omega_0$ for $g=0.01$~a.u.\ but only $0.59$~a.u.\ for $g=0.12$~a.u.\ (vertical bars in Figs.~\ref{fig:pecs}(c) and \ref{fig:pecs}(d)). In the same way, $|1,\xi_2\rangle$ is nearly resonant with $|0,\xi_4\rangle$ at large $R$ ($\epsilon_4-\epsilon_2=0.652$~a.u.\ at $R=10$~a.u.), so that the cavity becomes resonant with a transition of the dissociated fragment.

The two-level approximation fails quantitatively already for the lowest branches. See for example that branch $4$ for $g=0.01$~a.u. in Fig.~\ref{fig:pecs}(a) is already far off in that scale. For $g=0.12$~a.u., the TLS branches 1 and 2 deviate from the $N_{\text{PECs}}=10$ ones by up to $\sim10^{-2}$~a.u.  (Fig.~\ref{fig:pecs}(b) and Table~\ref{tab:pec_errors}); the error scales as $g^2$ and is almost independent of the DSE form, which shows that it originates in the truncation of the linear coupling, dominated by $\xi_3$. In the TLS the choice of the DSE form is much more important than in the MLS: the fDSE--tDSE difference of the lowest four TLS branches is more than an order of magnitude larger than in the MLS, because $\mathbb{M}^2$ truncated to two states poorly approximates $\mathbb{D}$.

\begin{table*}[t]
\caption{\label{tab:pec_errors}Accuracy of the polaritonic PECs: maximum absolute differences (a.u.) on $R\in[1,10]$~a.u., the region explored by the packet, for the branches $\alpha$ indicated, and minimum gap between branches 2 and 3 with its position. For pQED versus sepd, the differences are evaluated at the nodes of the sepd grid; at $g=0.01$~a.u.\ they reflect the different discretization of the electronic problem in the two calculations.}
\begin{ruledtabular}
\begin{tabular}{lll}
 & $g=0.01$~a.u. & $g=0.12$~a.u. \\
\hline
TLS vs MLS, fDSE ($\alpha=1,2$) & $4.4\times10^{-5}$, $1.1\times10^{-4}$ & $6.3\times10^{-3}$, $1.3\times10^{-2}$ \\
TLS vs MLS, tDSE ($\alpha=1,2$) & $5.1\times10^{-5}$, $1.1\times10^{-4}$ & $7.1\times10^{-3}$, $1.4\times10^{-2}$ \\
fDSE vs tDSE, MLS ($\alpha=1,2$) & $4.0\times10^{-6}$, $5.2\times10^{-6}$ & $7.5\times10^{-4}$, $8.4\times10^{-4}$ \\
fDSE vs tDSE, TLS ($\alpha\le4$) & $2.7\times10^{-4}$ & $3.6\times10^{-2}$ \\
pQED vs sepd, fDSE ($\alpha=1,2,3$) & $4.1\times10^{-4}$, $3.5\times10^{-4}$, $4.0\times10^{-4}$ & $6.2\times10^{-4}$, $6.0\times10^{-4}$, $1.3\times10^{-3}$ \\
pQED vs sepd, tDSE ($\alpha=1,2,3$) & $4.0\times10^{-4}$, $3.5\times10^{-4}$, $4.0\times10^{-4}$ & $8.3\times10^{-4}$, $8.2\times10^{-4}$, $8.0\times10^{-4}$ \\
Minimum gap $E_3-E_2$, fDSE / tDSE & $0.0176$ / $0.0176$ & $0.2075$ / $0.2073$ \\
Position of the minimum gap, fDSE and tDSE & $1.66$ & $1.67$ \\
\end{tabular}
\end{ruledtabular}
\end{table*}

The pQED and sepd curves coincide on the scale of Figs.~\ref{fig:pecs}(c) and \ref{fig:pecs}(d). Since the two calculations discretize the electronic problem differently, they differ already at $g=0.01$~a.u., by at most $4\times10^{-4}$~a.u.\ on $R\in[1,10]$~a.u.\ for $\alpha\le3$; this difference varies with $R$ and is the same for fDSE and tDSE (Table~\ref{tab:pec_errors}). At $g=0.12$~a.u.\ it grows only to $6\times10^{-4}$~a.u.\ for $\alpha=1,2$ and $1.3\times10^{-3}$~a.u.\ for $\alpha=3$ with fDSE, and to $8\times10^{-4}$~a.u.\ for $\alpha\le3$ with tDSE, i.e., to less than $1\%$ of the vacuum Rabi gap ($0.207$~a.u.). The fDSE and tDSE discrepancies are of the same order, so this test does not discriminate between the two forms of the DSE; it shows that reducing the dimensionality before (pQED) or after (sepd) the coupling to the cavity gives the same polaritonic PECs once both are converged.

The dynamics in the different approximation always have the same initial state given in Eq.~\eqref{eq:initial_state}. Figures~\ref{fig:pecs}(e) and \ref{fig:pecs}(f) show which branches the Franck--Condon state populates. Because the nuclear density is initially localized at the resonant crossing, $c^\alpha_{20}(R)$ (Sec.~\ref{sec:initial_state}) changes from 0 to 1 across the packet, and its weight is split almost equally between branches 2 and 3. Covering the initial state requires only a few branches for $g=0.01$~a.u., but several tens for $g=0.12$~a.u., a first indication that multiphoton and multistate components become relevant at the largest couplings.

\subsection{Nuclear dynamics and dissociation quenching}
\label{sec:res_densities}

\begin{figure*}[tbp]
  \centering
  \includegraphics[width=1.0\textwidth]{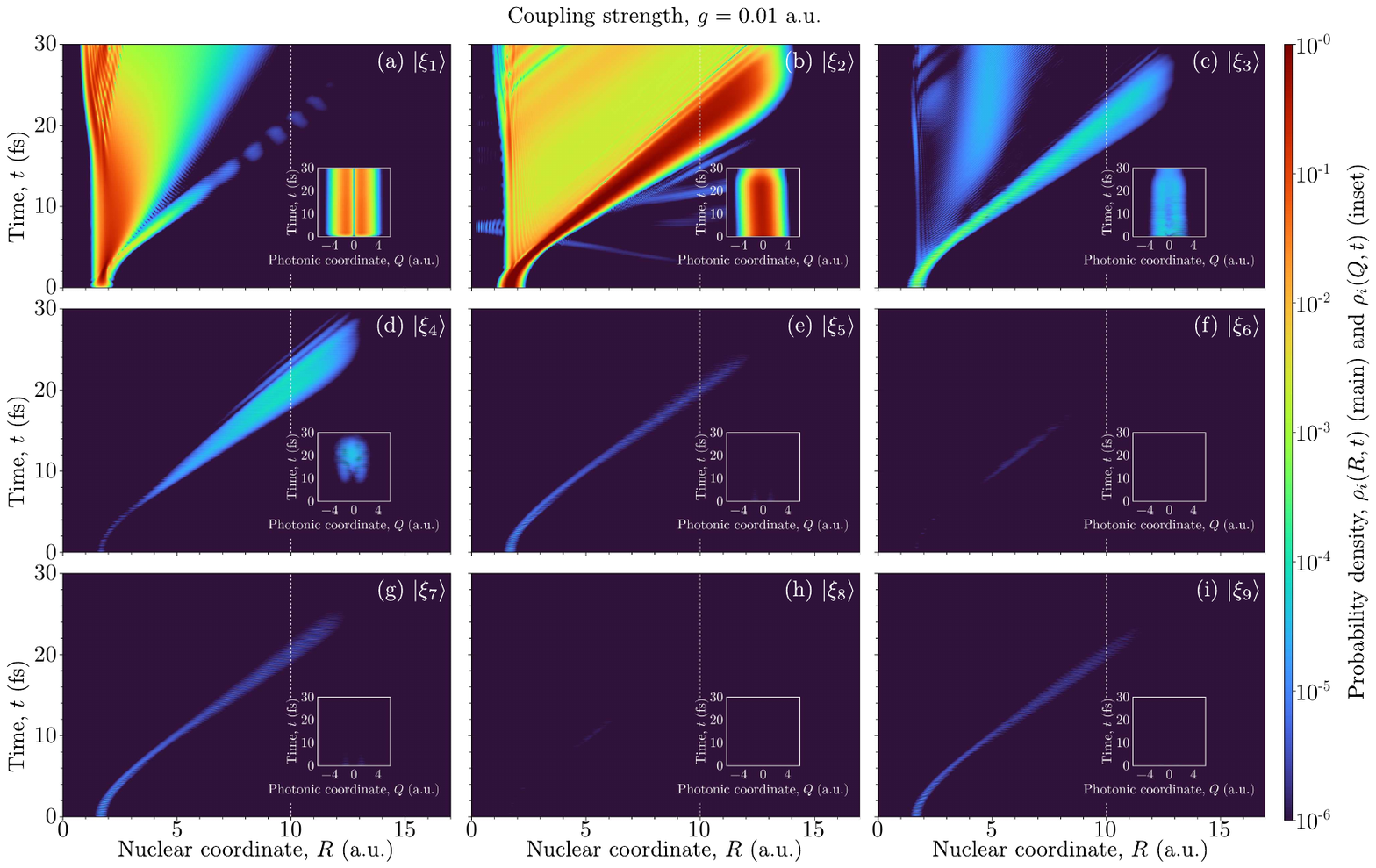}
  \caption{\label{fig:rho_g001}Nuclear probability density $\rho_i(R,t)$ (main panels) and photonic density $\rho_i(Q,t)$ (insets) projected on the electronic states (a) $\xi_1$ to (i) $\xi_9$, for $g=0.01$~a.u.\ (model \textsf{npd--fDSE--MLS}). Common logarithmic color scale from $10^{-6}$ to 1. The vertical dotted line marks the onset of the absorbing potential at $R_{\text{CAP}}=10$~a.u.}
\end{figure*}

\begin{figure*}[tbp]
  \centering
  \includegraphics[width=1.0\textwidth]{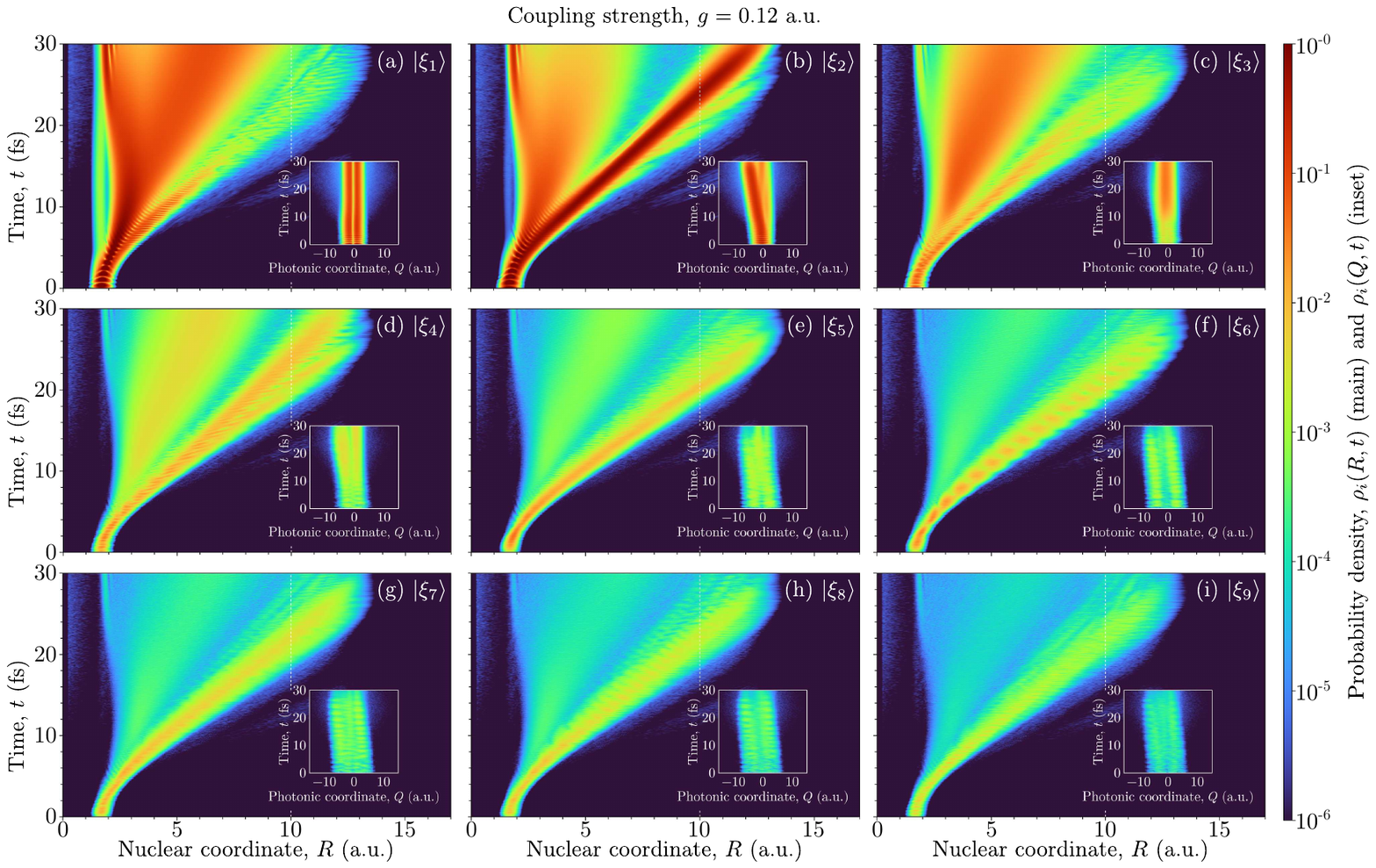}
  \caption{\label{fig:rho_g012}Same as Fig.~\ref{fig:rho_g001} for $g=0.12$~a.u. Note the different horizontal scale of the insets.}
\end{figure*}

Figures~\ref{fig:rho_g001} and \ref{fig:rho_g012} show the nuclear densities $\rho_i(R,t)$ and, in the insets, the photonic densities $\rho_i(Q,t)$ projected on the nine lowest electronic states, for $g=0.01$~a.u.\ and $g=0.12$~a.u. Table~\ref{tab:dynamics} collects the main quantities for all couplings.

For $g=0.01$~a.u.\ (Fig.~\ref{fig:rho_g001}), most of the packet dissociates on $\xi_2$ with the velocity of the bare excited PEC, as in the field-free dynamics~\cite{Mendez2025}. A fraction, however, is transferred within the first $\sim3$~fs to $\xi_1$, where it remains bound and oscillates in the potential well around $R\simeq2$~a.u.\ (Fig.~\ref{fig:rho_g001}(a)). The photonic density of this bound fraction has a node at all times, located at the displacement $Q_0=-\lambda\mu^{\text{BOA}}_{11}(R)/\omega_0\simeq-0.03$~a.u.\ produced by the PDM of $\xi_1$ (Eq.~\eqref{eq:completed_square}), which is indistinguishable from $Q=0$ on the scale of the inset. The bound fraction of the wave packet is then in the first excited state of the displaced mode, i.e., it carries one photon. This is the radiative transition $|0,\xi_2\rangle\to|1,\xi_1\rangle$ at the resonant crossing, through which the cavity converts part of the electronic excitation into a photon while the nuclei are still in the Franck--Condon region. Because $\epsilon_1$ is bound, the transferred population cannot dissociate: the cavity opens a radiative stabilization channel that quenches the dissociation. The interference fringes in $\rho_1$ and $\rho_2$ at short times reflect the coherent splitting of the packet over branches 2 and 3. The states $\xi_3,\dots,\xi_9$ are populated below $10^{-3}$.

\begin{table*}[t]
\caption{\label{tab:dynamics}Coupled nucleus--photon dynamics (\textsf{npd--fDSE--MLS}). $g$: coupling (a.u.); $p_{\xi_1}$ and $\langle R\rangle_1$: population of $\xi_1$ and mean nuclear distance (a.u.) in $\xi_1$ at $t=10$~fs, when $\mathcal{N}(t)=1$ to $10^{-4}$; {$p_{\xi_2}$: population of $\xi_2$ at $t=10$~fs;} $\mathcal{N}(30\,\text{fs})$: norm at $t=30$~fs (probability not yet absorbed at $R>10$~a.u.); $p_{\xi_{\ge3}}$: population of $\xi_3,\dots,\xi_{10}$ relative to $\mathcal{N}$ at 30~fs; $\overline{\langle\hat{n}_{\text{PZW}}\rangle}$ and $\overline{|\langle\hat{a}_{\text{PZW}}\rangle|}$: time averages over 0--18~fs, before the absorption at the CAP sets in, of the free-photon number and coherence; $\overline{\mathcal{G}^{(2)}_{\text{PZW}}}$: time average of $\mathcal{G}^{(2)}_{\text{PZW}}(0)$ over 5--18~fs, with the numerical tail of $p_n^{\text{phys}}$ replaced by a geometric series (Sec.~\ref{sec:res_statistics} {and Appendix~\ref{app:photon}}).  }
\begin{ruledtabular}
\begin{tabular}{ddddddddd}
\multicolumn{1}{c}{$g$} & \multicolumn{1}{c}{$p_{\xi_1}${ (10 fs)}} & \multicolumn{1}{c}{{$p_{\xi_2}$ (10 fs)}} & \multicolumn{1}{c}{$\langle R\rangle_1${ (10 fs)}} & \multicolumn{1}{c}{$\mathcal{N}(30\,\text{fs})$} & \multicolumn{1}{c}{$p_{\xi_{\ge3}}${ (30 fs)}} & \multicolumn{1}{c}{$\overline{\langle\hat{n}_{\text{PZW}}\rangle}$} & \multicolumn{1}{c}{$\overline{|\langle\hat{a}_{\text{PZW}}\rangle|}$} & \multicolumn{1}{c}{$\overline{\mathcal{G}^{(2)}_{\text{PZW}}}$} \\
\hline
0.01 & 0.137 & 0.863 & 2.30 & 0.145 & 0.000 & 0.13 & 0.017 & 0.011 \\
0.014 & 0.224 & 0.776 & 2.35 & 0.236 & 0.001 & 0.22 & 0.022 & 0.013 \\
0.02 & 0.332 & 0.668 & 2.41 & 0.344 & 0.002 & 0.32 & 0.022 & 0.015 \\
0.03 & 0.425 & 0.573 & 2.48 & 0.441 & 0.004 & 0.42 & 0.026 & 0.022 \\
0.04 & 0.474 & 0.523 & 2.56 & 0.496 & 0.010 & 0.47 & 0.053 & 0.036 \\
0.06 & 0.488 & 0.503 & 2.66 & 0.524 & 0.029 & 0.50 & 0.054 & 0.052 \\
0.08 & 0.481 & 0.492 & 2.80 & 0.558 & 0.078 & 0.52 & 0.052 & 0.051 \\
0.12 & 0.436 & 0.495 & 2.99 & 0.629 & 0.197 & 0.50 & 0.049 & 0.092 \\
\end{tabular}
\end{ruledtabular}
\end{table*}

\begin{figure*}[t]
  \centering
  \includegraphics{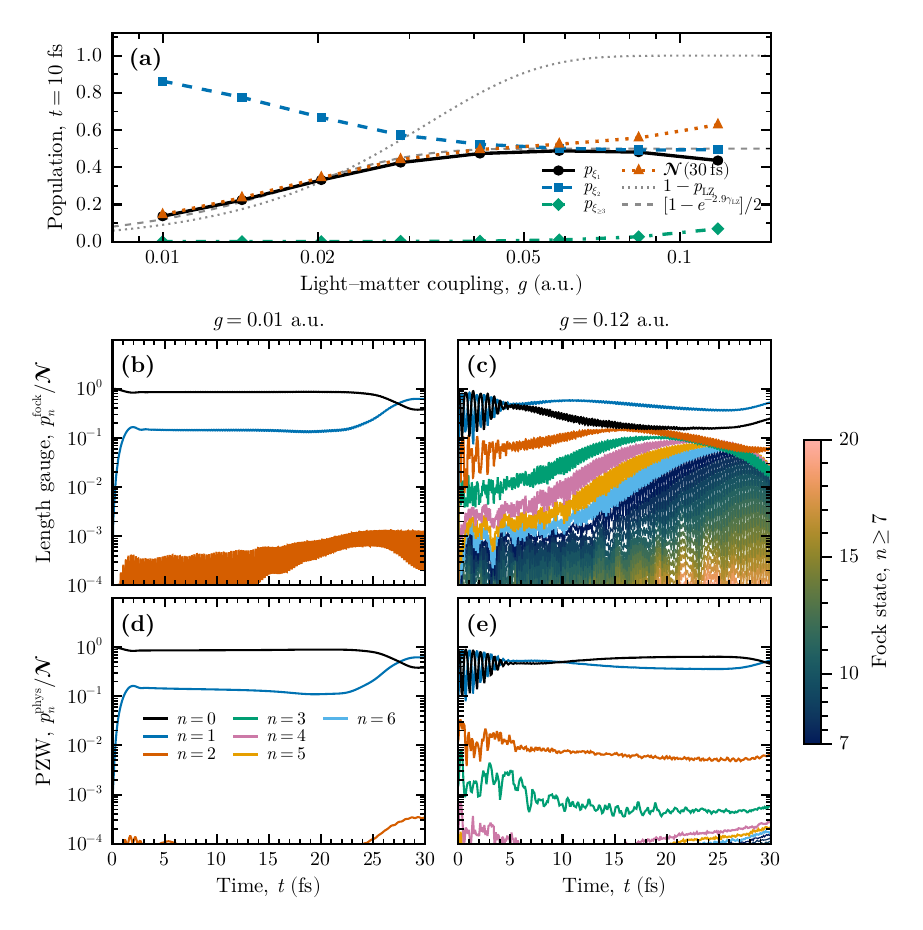}
  \caption{\label{fig:fock}Electronic populations and photon-number distributions of \textsf{npd--fDSE--MLS}. (a) Populations of $\xi_1$, $\xi_2$, and $\xi_3,\dots,\xi_{10}$ at $t=10$~fs, which add up to one because no probability has yet been absorbed, and norm $\mathcal{N}$ at $t=30$~fs, as functions of the coupling $g$ (Table~\ref{tab:dynamics}). Gray lines: Landau--Zener estimates $1-p_{\text{LZ}}=1-\exp(-\gamma_{\text{LZ}})$ (dotted) and $\frac12[1-\exp(-\zeta\gamma_{\text{LZ}})]$ with $\zeta=2.9$ (dashed), with $\gamma_{\text{LZ}}=0.095\,(g/0.01~\text{a.u.})^2$ (Sec.~\ref{sec:res_densities}). (b)--(e) Photon-number populations of the cavity mode, summed over the electronic states and conditioned on survival, for (b,d) $g=0.01$~a.u.\ and (c,e) $g=0.12$~a.u. (b,c) Length-gauge Fock states (eigenstates of $\hat{n}$), $p_n^{\text{fock}}(t)/\mathcal{N}(t)$, every 0.01~fs. (d,e) Free-photon number states (eigenstates of $\hat{n}_{\text{PZW}}$), $p_n^{\text{phys}}(t)/\mathcal{N}(t)$, every 0.1~fs. The states $n=0,\dots,6$ have individual colors (legend in panel (d)); those with $n=7,\dots,20$ are thin lines colored by $n$ (color bar). Only the $n\le20$ that exceed $10^{-4}$ at some time are shown.}
\end{figure*}

We quantify the efficiency of this channel by the population trapped in $\xi_1$ at $t=10$~fs, $p_{\xi_1}$, hereafter called the stabilization yield. It grows with the coupling (Table~\ref{tab:dynamics}), from $0.137$ for $g=0.01$~a.u.\ to $0.488$ for $g=0.06$~a.u., and the probability that has not reached the absorbing region after 30~fs increases from $0.145$ to $0.63$. At $t=10$~fs, no probability has yet been absorbed, and the population not trapped in $\xi_1$ is essentially on $\xi_2$, which dissociates ($p_{\xi_1}+p_{\xi_2}\geq0.99$ for $g\le0.06$~a.u.; Fig.~\ref{fig:fock}(a)). For $g\le0.06$~a.u., the norm at 30~fs is close to $p_{\xi_1}$: what survives is the stabilized fraction. The excess $\mathcal{N}(30\,\text{fs})-p_{\xi_1}$ at larger couplings is the population transferred to $\xi_{\ge3}$ and the part of $\xi_1$ that dissociates.

The saturation of $p_{\xi_1}$ near $1/2$ has a simple origin in the polaritonic picture. For small $g$ the gap at $R_{\text{eq}}$ is smaller than the energy spread of the packet and the width of the mixing region ($0.04$~a.u.\ for $g=0.01$~a.u.) is smaller than $\sigma_R$, so the packet crosses the resonance diabatically, following $|0,\xi_2\rangle$ toward dissociation. One can perform a Landau--Zener estimate~\cite{Zener1932}, $p_{\text{LZ}}=\exp(-\gamma_{\text{LZ}})$ for the probability of diabatic passage, with the adiabaticity parameter $\gamma_{\text{LZ}}=2\pi(g\mu^{\text{BOA}}_{12})^2/(\hbar v|\partial_R\Delta\epsilon_{12}|)$ and the asymptotic velocity of the bare packet, $v\simeq0.011$~a.u.\ (Sec.~\ref{sec:initial_state}). This gives $\gamma_{\text{LZ}}=0.095$ ($p_{\text{LZ}}\simeq0.91$) for $g=0.01$~a.u., so that only $1-p_{\text{LZ}}\simeq0.09$ of the packet follows the adiabatic branch, close to the computed $p_{\xi_1}=0.137$. Since $\gamma_{\text{LZ}}\propto g^2$, it reaches 1 at $g\simeq0.032$~a.u., and indeed $p_{\xi_1}$ rises from $0.22$ to $0.43$ between $g=0.014$ and $0.029$~a.u., the range of fastest change in Table~\ref{tab:dynamics} and Fig.~\ref{fig:fock}(a). For large $g$ the passage is adiabatic: each polaritonic branch is followed, and only the component on the {upper} branch{, $\alpha=3$,} with a weight close to $1/2$ remains bound. The two limits are interpolated by $p_{\xi_1}\simeq\frac12[1-\exp(-\zeta\gamma_{\text{LZ}})]$, which reproduces the computed yields within $0.03$ with $\zeta\simeq2.9$; $\zeta>1$ reflects that the packet starts at rest in the Franck--Condon region and crosses the resonance more slowly than its asymptotic velocity $v$. The mean distance $\langle R\rangle_1$ of the trapped packet increases from $2.3$ to $3.0$~a.u.\ with $g$, which reflects the cavity-induced deformation of the upper branch $\alpha=3$ (Fig.~\ref{fig:pecs}(f)).

For $g=0.12$~a.u.\ (Fig.~\ref{fig:rho_g012}), the dynamics is qualitatively richer. The transfer to $\xi_1$ is faster and accompanied by Rabi oscillations visible in $\rho_1$ and $\rho_2$ at short times, and the states $\xi_3,\dots,\xi_9$ acquire populations above $10^{-2}$; at $t=30$~fs they hold $19.7\%$ of the surviving probability (Table~\ref{tab:dynamics}). These states are reached through sequential photon exchange, starting with the near-resonant pair $|1,\xi_2\rangle\leftrightarrow|0,\xi_3\rangle$ identified in Sec.~\ref{sec:res_dipoles}. Their packets dissociate on the corresponding bare PECs, and part of the population of $\xi_1$ also dissociates: at 30~fs, $\langle R\rangle_1=5.2\pm2.2$~a.u., compared with $1.8\pm0.6$~a.u.\ for $g=0.01$~a.u. The nonmonotonic behavior of $p_{\xi_1}$ beyond $g\approx0.06$~a.u.\ is therefore due to the redistribution of the population over states outside the TLS. Since $p_{\xi_{\ge3}}$ exceeds $7\%$ for $g\gtrsim0.08$~a.u., the TLS is not adequate in this range, in agreement with the static analysis of Sec.~\ref{sec:res_pecs}.

The photonic densities in the insets are displaced from $Q=0$ (unlike models that do not include PDMs, see Fig.~S1 of the Supplemental Material~\cite{SM}), and the displacement depends on the electronic state and grows with $g$: for $g=0.12$~a.u., $\langle Q\rangle_2$ reaches $\sim-3$~a.u.\ at 30~fs, while $\langle Q\rangle_1$ stays close to $0$ (Fig.~\ref{fig:rho_g012}, insets). This is the shift $Q_0=-\lambda\mu^{\text{BOA}}_{ii}(R)/\omega_0$ of the minimum of Eq.~\eqref{eq:completed_square} for a molecule with PDMs: in the length gauge, $Q$ is a displacement-field coordinate, and its distribution follows the {molecular dipole}, which grows as the fragments separate. The displacement does not represent free photons, a point developed in Sec.~\ref{sec:res_coherence}.

\subsection{Full electron--nucleus--photon dynamics}
\label{sec:res_full}

\begin{figure*}[t]
  \centering
  \includegraphics[width=0.92\textwidth]{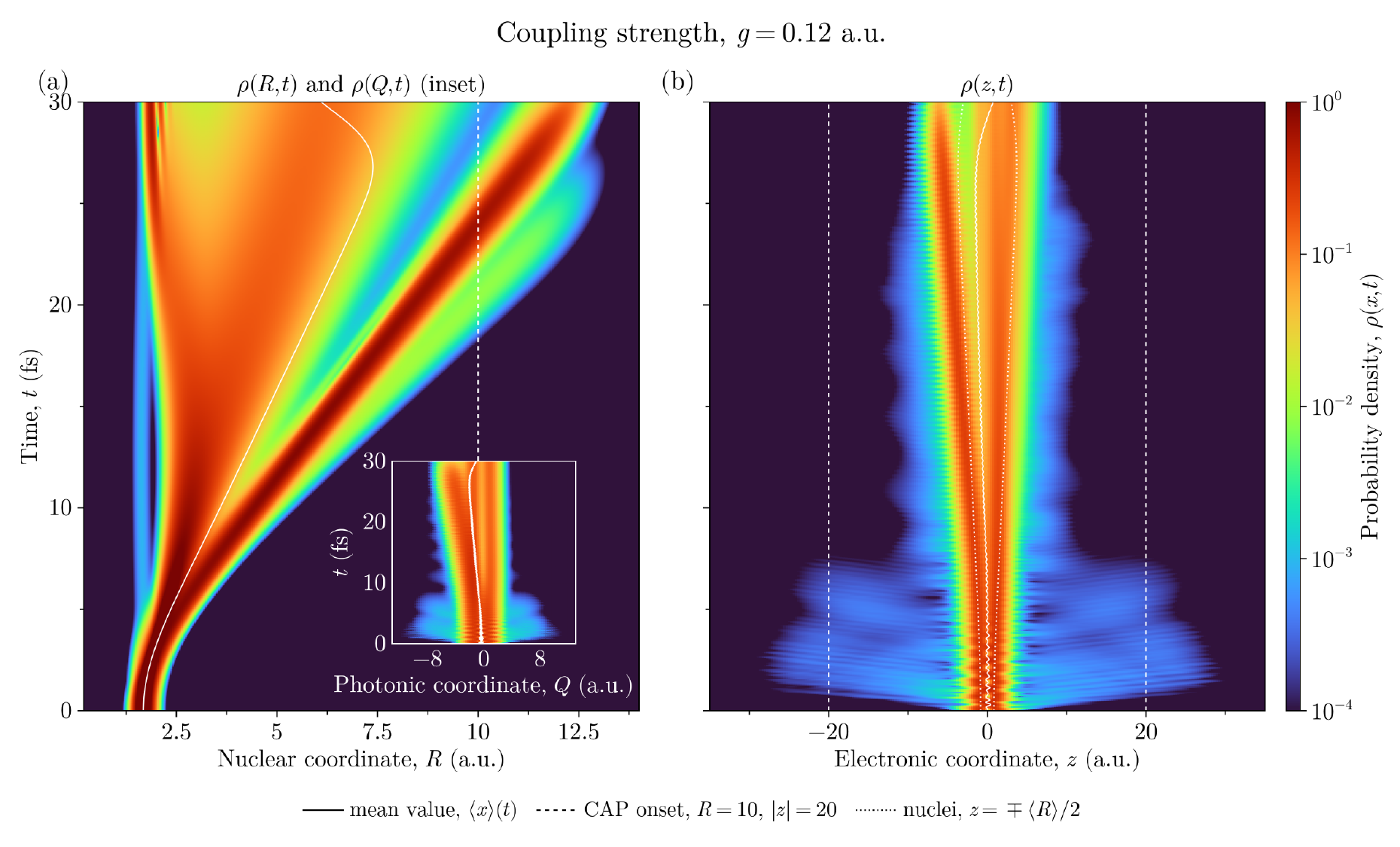}
  \caption{\label{fig:rho_full}Reduced one-dimensional densities of the full electron--nucleus--photon dynamics (model \textsf{enpd}) for $g=0.12$~a.u., as functions of time, each obtained by integrating $|\Psi(R,z,Q,t)|^2$ over the other two coordinates. (a) Nuclear density $\rho(R,t)$ (main panel) and photonic density $\rho(Q,t)$ (inset). (b) Electronic density $\rho(z,t)$, with $z$ measured from the internuclear midpoint. Common logarithmic color scale from $10^{-4}$ to 1. White lines: mean value $\langle x\rangle(t)$ (solid), onset of the absorbing potentials at $R=10$~a.u.\ and $|z|=20$~a.u.\ (dashed), and nuclear positions $z=\mp\langle R\rangle/2$ (dotted).}
\end{figure*}

The previously discussed \textsf{npd--fDSE--MLS} results rest on three approximations: the truncation to ten electronic states, the neglect of the bare nonadiabatic couplings $d_{ij}$, and a discretized continuum. In the \textsf{enpd} model (Sec.~\ref{sec:methods_full}) all three are absent. Figure~\ref{fig:rho_full} shows its reduced densities. The nuclear density (Fig.~\ref{fig:rho_full}(a)) reproduces the two components identified in Fig.~\ref{fig:rho_g012}: a packet that remains in the Franck--Condon region, oscillating around $R\simeq2$~a.u., and a dissociating jet that reaches the absorbing region at $t\simeq20$~fs, together with the slower fraction that spreads between $R\simeq3$ and $7$~a.u.\ at late times. The photonic density (inset) shows the same two lobes, displaced toward negative $Q$, as the projection on $\xi_1$ in Fig.~\ref{fig:rho_g012}(a). The norm at 30~fs is {$\mathcal{N}(30\,\text{fs})=0.624$, compared with $0.629$} in \textsf{npd--fDSE--MLS} (Table~\ref{tab:dynamics}): the dissociation quenching is therefore not an artifact of the electronic truncation. Most of the remaining probability, $0.350$, is absorbed at large $R$, and only $p_z=0.026$ by the electronic absorbing potential, almost all of it during the first $\sim7$~fs.

The electronic density (Fig.~\ref{fig:rho_full}(b)) splits into two branches that follow the nuclei, $z\simeq\mp\langle R\rangle/2$, as the molecule dissociates: the electron localizes on one center or the other, which is the charge-transfer character responsible for the linear growth of the PDMs (Sec.~\ref{sec:res_dipoles}). A weak density, of order $10^{-4}$--$10^{-3}$, reaches $|z|>20$~a.u.\ during the first $\sim7$~fs; it corresponds to electron emission into the continuum and accounts for $p_z$. This ionization channel, which the ten-state expansion can only mimic through its pseudocontinuum states, removes a small fraction of the probability and does not modify the branching between the bound and dissociative channels.

\subsection{Photon-number distribution}
\label{sec:res_fock}

As discussed in Sec.~\ref{sec:pf_hamiltonian}, the length-gauge photon number $\langle\hat{n}\rangle$ is often interpreted as the number of photons in the cavity, although only $\langle\hat{n}_{\text{PZW}}\rangle$ (Eq.~\eqref{eq:physical_photon}) counts free photons. Here we compare the two distributions and relate them to the dissociation. Figure~\ref{fig:fock} compares the populations of the length-gauge Fock states, $p_n^{\text{fock}}(t)/\mathcal{N}(t)$ (panels (b) and (c)), with those of the free-photon number states, $p_n^{\text{phys}}(t)/\mathcal{N}(t)$ (panels (d) and (e)). The former are the populations of the photon-number basis in which the polaritonic states of Sec.~\ref{sec:methods_pecs} are expanded; they count quanta of the displacement field, which include the static field bound to the molecular polarization. For $g=0.01$~a.u., both distributions nearly coincide. The population of the one-photon state rises within $\sim2$~fs to $0.14$ and stays nearly constant until the dissociating fragments reach the absorbing potential; this is the photon emitted in the radiative stabilization. Its probability, $p_1^{\text{phys}}/\mathcal{N}=0.137$ at 10~fs, coincides with the population $p_{\xi_1}$ of $\xi_1$ in Table~\ref{tab:dynamics}: each stabilized molecule has emitted exactly one free photon, and the state is close to the superposition $\sqrt{1-p_{\xi_1}}\,|0,\xi_2\rangle+\sqrt{p_{\xi_1}}\,|1,\xi_1\rangle$ (with the corresponding dissociating and bound nuclear packets). The photon number and the electronic state are thus entangled: finding one photon in the cavity implies that the molecule is bound in $\xi_1$, and finding none implies that it dissociates on $\xi_2$ (Sec.~\ref{sec:res_entanglement}). After $\sim20$~fs, absorption removes the dissociating $|0,\xi_2\rangle$ component, and the conditioned population of $|1\rangle$ grows toward unity: the surviving molecule is, to a good approximation, in $|1,\xi_1\rangle$.

For $g=0.12$~a.u., the two descriptions differ qualitatively. The length-gauge distribution extends beyond $n=20$, and the populations with $n\ge2$ build up progressively as the PDMs of the dissociating fragments grow and the length-gauge vacuum becomes increasingly displaced. In the free-photon distribution, after the Rabi oscillations of the first 5~fs, only $n=0$ and $1$ are appreciably populated, with weights close to $1/2$, and $\sum_{n\ge2}p_n^{\text{phys}}/\mathcal{N}$ stays of order $10^{-2}$: the cavity contains essentially the single photon emitted in the stabilization, and the multiphoton content of the length-gauge distribution is the polarization field. The initial state, which is the length-gauge vacuum, already contains free photons, $\langle\hat{n}_{\text{PZW}}\rangle(0)=E_{\text{q}}/\hbar\omega_0$ (Eq.~\eqref{eq:quench_energy}), that is, {$0.12$ free photons on average for the largest coupling, $g=0.12$~a.u.}

\subsection{Coherence and polarization field}
\label{sec:res_coherence}

The length-gauge coherence grows from $|\langle\hat{a}\rangle|\simeq10^{-2}$ at short times to $\sim0.1$ for $g=0.01$~a.u.\ and to $\sim1$ for $g=0.12$~a.u.\ (gray lines in Fig.~\ref{fig:photon}(a)); its real part is negative and grows in magnitude with time (Table~\ref{tab:static}), and an oscillation with the photon period $2\pi/\omega_0=0.236$~fs is superimposed. Because the PDMs break the photon-parity symmetry discussed in Sec.~\ref{sec:Rabi_model}, $\langle\hat{a}\rangle\neq0$ is allowed here. Its physical content follows from Eq.~\eqref{eq:physical_photon}: if the molecular dipole varies slowly on the scale of the photon period, so that the field adiabatically follows it, the free-photon amplitude $\langle\hat{a}_{\text{PZW}}\rangle$ vanishes and
\begin{equation}\label{eq:static_estimate}
  \langle\hat{a}\rangle\simeq-\frac{g}{\hbar\omega_0}\langle\hat{\mu}_\epsilon\rangle\simeq-\frac{g}{\hbar\omega_0}\sum_i\frac{p_{\xi_i}}{\mathcal{N}}\,\mu^{\text{BOA}}_{ii}\big(\langle R\rangle_i\big),
\end{equation}
where the second step neglects the electronic coherences and the width of the packets. Since $\hat{\mu}_\epsilon$ is Hermitian, Eq.~\eqref{eq:physical_photon} gives $\text{Im}\,\langle\hat{a}\rangle=\text{Im}\,\langle\hat{a}_{\text{PZW}}\rangle$ exactly: the static field displaces only $\text{Re}\,\langle\hat{a}\rangle\propto\langle\hat{Q}\rangle$, whereas $\text{Im}\,\langle\hat{a}\rangle\propto\langle\hat{P}\rangle$ is part of the free-photon oscillation. The comparison is therefore made for the real part. This static estimate, evaluated with the computed populations and mean distances, reproduces $\text{Re}\,\langle\hat{a}\rangle$ averaged over one photon period to within $10\%$ for $g\le0.06$~a.u.\ and $17\%$ for all couplings, for $2\le t\le30$~fs, with median deviations of 1--5\% (Appendix~\ref{app:photon}); for $g=0.12$~a.u.\ at $t=20$~fs, for instance, Eq.~\eqref{eq:static_estimate} gives $-1.00$ against the computed $-0.99$.

The coherence is therefore dominated by the static polarization field bound to the PDMs, and grows with time because $\mu^{\text{BOA}}_{22}\propto R$ for the dissociating fragments. The direct evaluation of the free-photon amplitude confirms this picture (Fig.~\ref{fig:photon}(a)): for $t\ge5$~fs, $|\langle\hat{a}_{\text{PZW}}\rangle|$ remains below $\sim0.1$ for all couplings, with time averages over 0--18~fs from $0.017$ to $0.054$ (Table~\ref{tab:dynamics}), about an order of magnitude below $|\langle\hat{a}\rangle|$ (gray lines in the figure), and its trajectory in the quadrature plane is a rotation with the photon period around the origin {(Fig.~\ref{fig:photon}(b))}, without the growing displacement of $\langle\hat{a}\rangle$. Moreover, for a charged molecule the absolute value of $\langle\hat{a}\rangle$ depends on the dipole origin, since a shift $\hat{\mu}_\epsilon\to\hat{\mu}_\epsilon+\delta\mu$ changes it by $-(g/\hbar\omega_0)\delta\mu$, whereas $\langle\hat{a}_{\text{PZW}}\rangle$ is origin independent.
In particular, $\langle\hat{a}\rangle\neq0$ does not signal a symmetry-breaking transition: the model has a finite number of degrees of freedom, and the photon-parity symmetry is broken explicitly by the PDMs in the Hamiltonian. Without PDMs (Sec.~S4 of the Supplemental Material~\cite{SM}), both $\langle\hat{a}\rangle$ and $\langle\hat{a}_{\text{PZW}}\rangle$ vanish identically for the initial state of Eq.~\eqref{eq:initial_state}. A spontaneous breaking of a discrete symmetry, as in the superradiant transition of the Dicke and Rabi models~\cite{Hepp1973,ashhab2013superradiance,hwang2015quantum}, requires either a thermodynamic limit or a vanishing frequency ratio $\omega_0/\Delta\epsilon\to0$~\cite{hwang2015quantum}, and for atomic ensembles in the dipole approximation it is excluded when the $A^2$ or DSE term is retained, as a consequence of the TRK sum rule~\cite{Rzazewski1975,Bialynicki1979,Nataf2010,Andolina2019}. Consistently, neither the free-photon observables of Fig.~\ref{fig:averages} nor the weighted entanglement entropy (Sec.~\ref{sec:res_entanglement}) show a critical coupling in the range studied.

\subsection{Intracavity photon statistics}
\label{sec:res_statistics}

\begin{figure*}[tbp]
  \centering
  \includegraphics[width=1.0\textwidth]{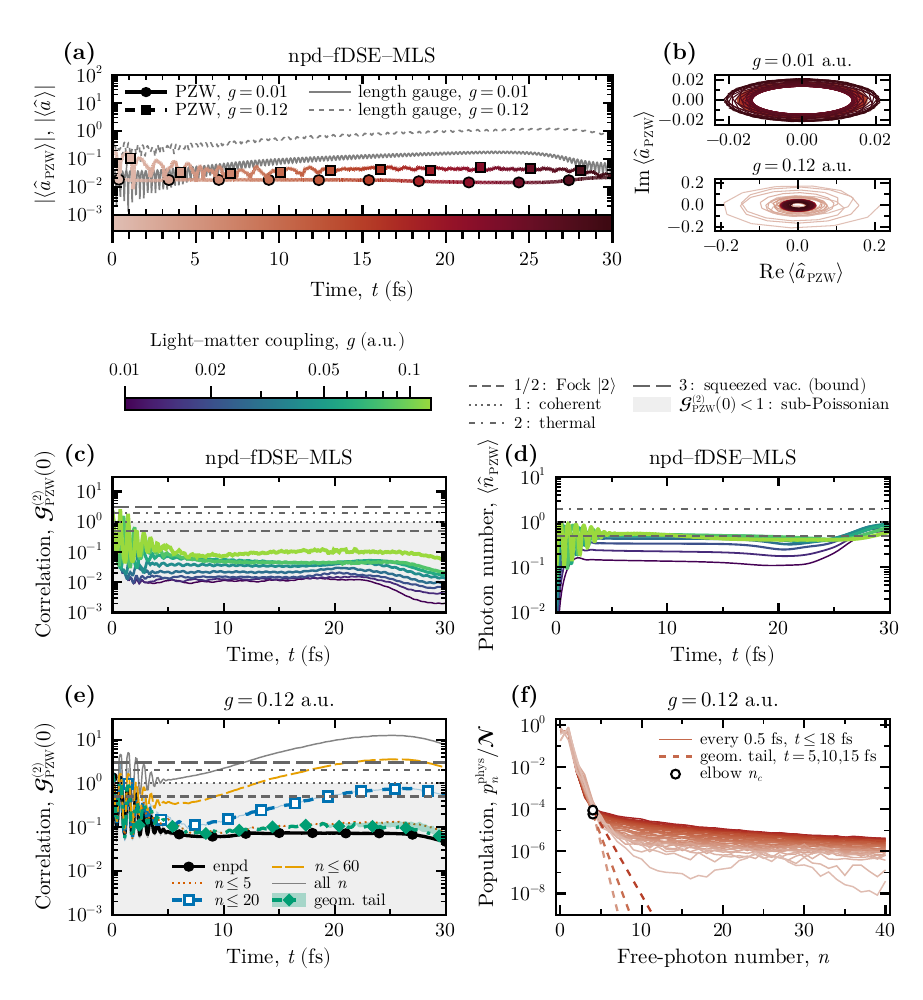}
  \caption{\label{fig:photon}Free-photon field, conditioned on survival. (a) Free-photon coherence $|\langle\hat{a}_{\text{PZW}}\rangle|(t)$ for $g=0.01$~a.u.\ (solid line, circles) and $g=0.12$~a.u.\ (dashed line, squares), colored by time (color bar below); thin gray lines: length-gauge coherence $|\langle\hat{a}\rangle|(t)$. (b) Corresponding trajectories in the quadrature plane ($g=0.01$~a.u., top; $g=0.12$~a.u., bottom). (c) $\mathcal{G}^{(2)}_{\text{PZW}}(0)$ with the geometric tail of Eq.~\eqref{eq:G2geo} and (d) $\langle\hat{n}_{\text{PZW}}\rangle(t)$ for the eight couplings (color and thickness increase with $g$). Panels (a)--(d): \textsf{npd--fDSE--MLS}. (e) $\mathcal{G}^{(2)}_{\text{PZW}}(0)$ for $g=0.12$~a.u.\ in \textsf{enpd} and in \textsf{npd--fDSE--MLS} with all $n$, with cutoffs at $n\le5$, 20, and 60, and with the geometric tail (band: elbow threshold between 0.3 and 0.5); running averages over one photon period, with instantaneous values as light lines. (f) $p_n^{\text{phys}}/\mathcal{N}$ of \textsf{npd--fDSE--MLS} for $g=0.12$~a.u.\ up to 18~fs, colored by time; at 5, 10, and 15~fs, open circles mark the elbow $n_c=4$ and dashed lines the geometric continuations. Gray reference lines (Sec.~\ref{sec:observables}): $\mathcal{G}^{(2)}_{\text{PZW}}(0)=1/2$ (Fock state $|2\rangle$), 1 (coherent), 2 (thermal), and 3 (squeezed-vacuum lower bound) in (c) and (e), and $\langle\hat{n}_{\text{PZW}}\rangle=1/2$, 1, and 2 in (d). Shaded region: sub-Poissonian statistics.}
\end{figure*}

Figures~\ref{fig:photon}(c) and \ref{fig:photon}(d) show $\mathcal{G}^{(2)}_{\text{PZW}}(0)$ and $\langle\hat{n}_{\text{PZW}}\rangle$  as a function of time for the eight computed couplings. The mean free-photon number rises within $\sim2$~fs and stays nearly constant until the fragments reach the absorbing potential. Its time average over 0--18~fs, before the absorption at the CAP sets in, grows from $0.13$ for $g=0.01$~a.u.\ to $\simeq0.5$ for $g\ge0.04$~a.u.\ and then saturates as the stabilization yield does (Table~\ref{tab:dynamics}). At 10~fs, $\langle\hat{n}_{\text{PZW}}\rangle\simeq0.14$ for $g=0.01$~a.u., equal to $p_{\xi_1}$. While for $g=0.12$~a.u. $\langle\hat{n}_{\text{PZW}}\rangle\simeq0.5$ which is above $p_{\xi_1}=0.44$, the excess here coming from the initial free-photon content $E_{\text{q}}/\hbar\omega_0$ and from the sequential photon exchange with higher states.

\begin{figure*}[t]
  \centering
  \includegraphics{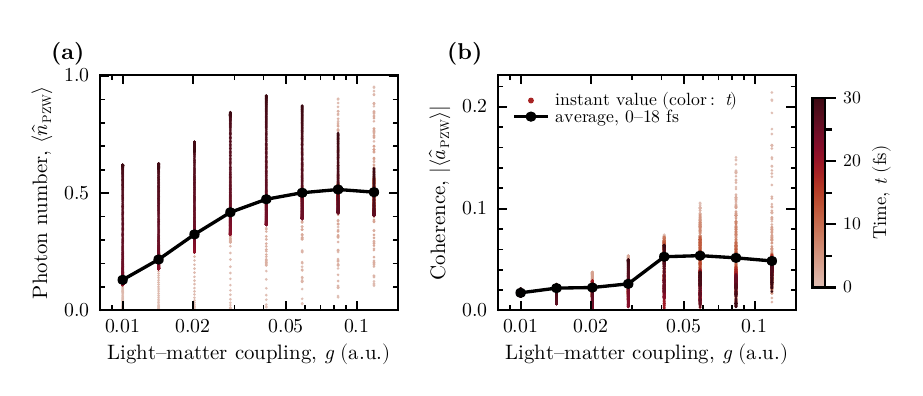}
  \caption{\label{fig:averages}Free-photon observables of \textsf{npd--fDSE--MLS} as functions of the coupling $g$. (a) Mean free-photon number $\langle\hat{n}_{\text{PZW}}\rangle$ and (b) free-photon coherence $|\langle\hat{a}_{\text{PZW}}\rangle|$, conditioned on survival: dots are instantaneous values every 0.05~fs over 0--30~fs, colored by time (color bar), and the solid line with circles is the time average over 0--18~fs, before the absorption at the CAP sets in (Table~\ref{tab:dynamics}). At later times (darkest dots), the surviving fraction is enriched in the bound population of $\xi_1$.}
\end{figure*}

The length-gauge photon number $\langle\hat{n}\rangle$ contains in addition the term $(g/\hbar\omega_0)^2\langle\hat{\mu}_\epsilon^2\rangle$ associated with the polarization (Eq.~\eqref{eq:physical_photon}), which grows with $g$ and with time, as the PDMs of the dissociating fragments grow linearly with $R$ (Sec.~\ref{sec:res_dipoles}). The mean free-photon number is insensitive to the electronic description: for $g=0.12$~a.u., the \textsf{enpd} and \textsf{npd--fDSE--MLS} values of $\langle\hat{n}_{\text{PZW}}\rangle$, averaged over one photon period, agree to within 6\% for $5\le t\le18$~fs. The instantaneous values of $\langle\hat{n}_{\text{PZW}}\rangle$ and $|\langle\hat{a}_{\text{PZW}}\rangle|$ (Figs.~\ref{fig:averages}(a) and \ref{fig:averages}(b)) spread around their time averages over 0--18~fs  which increase with the coupling up to $g\simeq0.04$~a.u.\ and then saturate (Table~\ref{tab:dynamics}).

The free-photon statistics is sub-Poissonian. For $g\le0.04$~a.u., $\mathcal{G}^{(2)}_{\text{PZW}}(0)$ is of order $10^{-2}$ (time averages over 5--18~fs from $0.011$ to $0.036$): the field is dominated by the single photon emitted in the radiative transition, heralded by the electronic state $\xi_1$, and a mixture of $|0\rangle$ and $|1\rangle$ has $\mathcal{G}^{(2)}_{\text{PZW}}(0)=0$.

For larger couplings, $\mathcal{G}^{(2)}_{\text{PZW}}(0)$  behavior is very sensitive to the population distribution. In the model \textsf{npd--fDSE--MLS}, evaluated with the complete distribution, $\mathcal{G}^{(2)}_{\text{PZW}}(0)$ is dominated by a small, nearly flat tail of $p_n^{\text{phys}}$, of order $10^{-6}$ at 10~fs and $10^{-5}$ at 25~fs for $g=0.12$~a.u.\ This small population extends up to the  maximum number used in  the photon basis (Fig.~\ref{fig:photon}(f)). The tail is negligible in the norm and in $\langle\hat{n}_{\text{PZW}}\rangle$, but each state enters the numerator with weight $n(n-1)$: from 10~fs on, states with $n>20$ contribute 76--81\% of it, and $\mathcal{G}^{(2)}_{\text{PZW}}(0)$ ranges between 2 and 12, against 0.06--0.07 in \textsf{enpd} over the same interval. Three observations show that the tail is not physical. A free photon costs $\hbar\omega_0=0.645$~a.u., while the energy available above the minimum of $\epsilon_1$ is $0.72\pm0.05$~a.u., so that only one free photon fits and components with $n\ge2$ can only be virtual and small. The tail is already present in the length-gauge distribution $p_n^{\text{fock}}$, which is read from the wave function without the PZW transformation, so it is not created by $\hat{U}_{\text{PZW}}$, whose field displacements, $g|\mu_\epsilon|/\hbar\omega_0$, are $\lesssim2$. Finally, for $g=0.014$~a.u., the populations with $n\ge6$ change by factors of 2 to 10 between propagations with different integrator tolerances, whereas those with $n\le4$ change by less than 10\%.

These observations lead us to propose a cutoff of the distribution at some $n_{\text{cut}}$ which removes most of the tail. The result, however, grows monotonically with $n_{\text{cut}}$ (Fig.~\ref{fig:photon}(e)), and no cutoff follows from the dynamics. Two parts of the distribution are clearly separated using the ratios $p_{n+1}^{\text{phys}}/p_n^{\text{phys}}$. The physical decay is below 0.3 per photon ($q=0.03$--$0.21$ at 5--15~fs for $g=0.12$~a.u.), whereas in the tail the ratio is $0.6$--$1$ (Fig.~\ref{fig:photon}(f)). We therefore keep the computed populations up to the elbow $n_c$ at which the physical decay ends and replace the tail by a geometric series that continues it, which gives $\mathcal{G}^{(2)}_{\text{PZW}}(0)$ in closed form (Appendix~\ref{app:photon}).

The geometric value is of the same order as that of the \textsf{enpd} propagation, which gives  $\mathcal{G}^{(2)}_{\text{PZW}}(0)\approx 0.07$ on average, (Fig.~\ref{fig:photon}(e)).  Hence, the free-photon statistics is strongly sub-Poissonian for all couplings. Nonclassicality criteria based only on the lowest photon-number probabilities, such as that of Klyshko~\cite{KLYSHKO19967},  could provide an alternative that is less sensitive to the tail.

Note that $\mathcal{G}^{(2)}_{\text{PZW}}(0)$ characterizes the equal-time photon-number distribution of the intracavity free field, not the statistics of the emitted photons. Here $\mathcal{G}^{(2)}_{\text{PZW}}(0)<1$ indicates sub-Poissonian statistics, whereas antibunching refers to delayed coincidences, $\mathcal{G}^{(2)}(\tau)>\mathcal{G}^{(2)}(0)$. Moreover, at ultrastrong coupling the free-photon number also counts virtual photons that dress the molecule and would not leave a lossy cavity~\cite{Ridolfo2012}. Even at $t=0$, before any radiative transition, $p_1^{\text{phys}}=0.093$ for $g=0.12$~a.u., because the length-gauge vacuum of Eq.~\eqref{eq:initial_state} is displaced from the free-field vacuum (Eq.~\eqref{eq:physical_photon}); these free photons carry the quench energy $E_{\text{q}}$ (Sec.~\ref{sec:initial_state}).

These intracavity quantities are, in principle, accessible through nondemolition measurements of the photon number~\cite{Gleyzes2007,Guerlin2007,Deleglise2008} or intracavity tomography~\cite{Choi2025}. A comparison with emission experiments, which probe the output field through input--output theory~\cite{WallsMilburn2008,gerry_introductory_2004}, would require the inclusion of cavity losses.

\subsection{Nucleus--photon and light--matter entanglement}
\label{sec:res_entanglement}

\begin{figure*}[t]
  \centering
  \includegraphics[width=\textwidth]{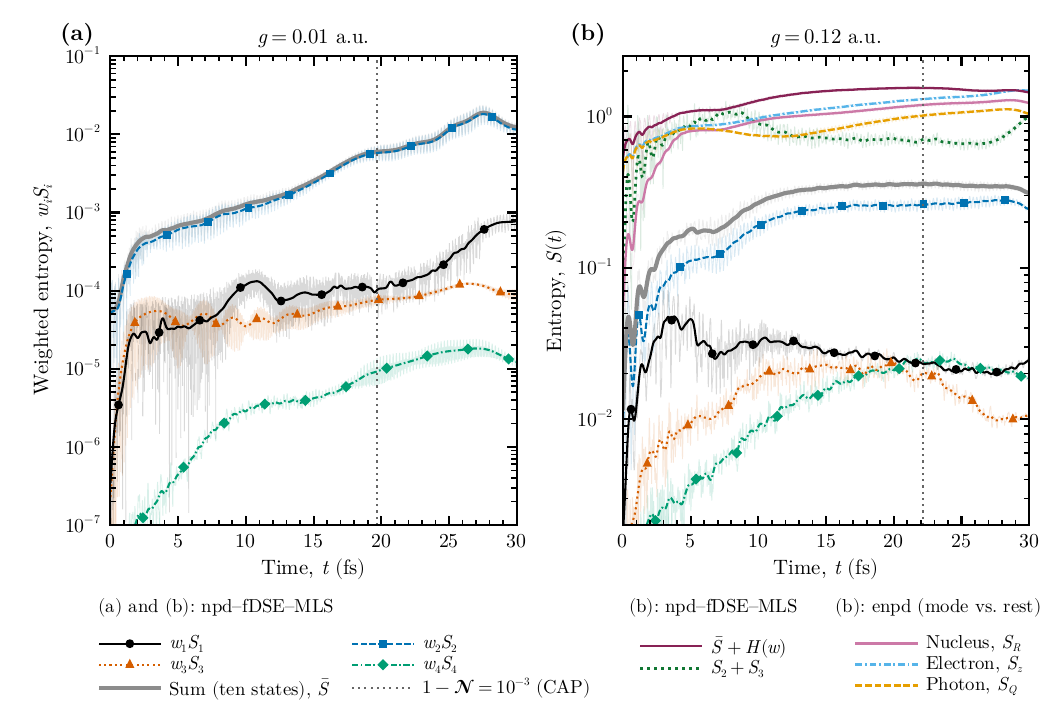}
  \caption{\label{fig:entropy}Entanglement entropies decomposed by electronic state, conditioned on survival, on a logarithmic scale. Thick lines are the instantaneous values filtered with a Gaussian of width $\sigma=0.15$~fs, and the thin light lines behind them are the instantaneous values. (a) $g=0.01$~a.u., \textsf{npd--fDSE--MLS}: contributions $w_iS_i$, with $w_i=p_{\xi_i}/\mathcal{N}$ and $S_i$ the $R$--$Q$ entanglement entropy within $\xi_i$ (Eq.~\eqref{eq:vn_entropy}), of the states $\xi_1,\dots,\xi_4$, and the weighted entropy over the ten states, $\bar{S}=\sum_iw_iS_i$ (thick gray line). (b) $g=0.12$~a.u.: the same quantities of \textsf{npd--fDSE--MLS}, the upper bound $\bar{S}+H(w)$ of the photon entropy, with $H(w)=-\sum_iw_i\ln w_i$, and $S_2+S_3$ (unweighted); and the entropy of each mode with the other two in \textsf{enpd}, $S_R$, $S_z$, and $S_Q$. Negative numerical artifacts of $S_i$, which appear when $p_{\xi_i}\lesssim10^{-7}$, are set to zero. The vertical dotted line marks $1-\mathcal{N}=10^{-3}$ in \textsf{npd--fDSE--MLS} (19.7~fs for $g=0.01$~a.u.\ and 22.1~fs for $g=0.12$~a.u.); at later times, the results are affected by the absorption at the CAP.}
\end{figure*}

In this section we will discuss the entanglement between constituents in the full \textsf{enpd} and relate it to the nuclei-photon entanglement computed for each electronic level in the multilevel approach. Figure~\ref{fig:entropy} shows the nucleus--photon entanglement by electronic state: the contributions $w_iS_i$ of the four lowest states and the weighted entropy $\bar{S}$ (Sec.~\ref{sec:observables}) for $g=0.01$ and $0.12$~a.u.

The lower channel $\xi_1$ represents a bound vibrational state of the molecule. In this case, the nucleus and the photon are nearly unentangled ( at 10~fs $S_1\simeq6\times10^{-4}$ for $g=0.01$~a.u.\ and $0.09$ for $g=0.12$~a.u.\ ), since the trapped fraction is close to a product of a vibrational packet and a one-photon state. On the other hand, the dissociative channel $\xi_2$ gives the largest contribution over the studied coupling values: over the time range  1--18~fs, $w_2S_2$ is 81--96\% of $\bar{S}$ for $g=0.01$~a.u.\ and 51--74\% for $g=0.12$~a.u. Two mechanisms correlate $Q$ with $R$ in this channel: the photon exchange at the resonant crossing, which is confined to the Franck--Condon region, and the $R$-dependent displacement $\beta_2(R)=-g\mu^{\text{BOA}}_{22}(R)/\hbar\omega_0$ of the polarization field along the dissociating packet.

Higher states have larger conditional entropies ($S_i\sim0.5$--$2$, not shown), because they are reached through multiphoton processes localized in space and time, but they are very weakly populated and enter $\bar{S}$ with weights $w_i\lesssim10^{-1}$ for $g=0.12$~a.u.\ and times between $10--18$~fs. Specifically,  $\xi_4,\dots,\xi_{10}$ together contribute only 10--13\% of $\bar{S}$.

Regarding the coupling strength dependence, between 5 and 18~fs, $\bar{S}$ is about 70 to 270 times smaller for $g=0.01$ than for $g=0.12$~a.u.\ (filtered curves). The average over this interval, which precedes the absorption at the CAP for all couplings, grows monotonically with $g$, from $1.9\times10^{-3}$ to $0.28$.

Note that all entropies oscillate with a high frequency that matches that of the rotation of the free-photon coherence $\langle\hat{a}_{\text{PZW}}\rangle$ in the quadrature plane (Fig.~\ref{fig:photon}{(b)}) This frequency is modified from $\simeq\omega_0$ for $g\le0.06$~a.u.\ to $\simeq0.87\,\omega_0$ for $g=0.12$~a.u., in both \textsf{npd--fDSE--MLS} and \textsf{enpd}. The oscillation is removed from the curves by the filter described in {Appendix~\ref{app:photon}}.

In the \textsf{enpd} model, correlations between all particles can be characterized. One can then relate the weighted entropy $\bar{S}$ or the electronic channel entropies $S_i$ to the three bipartite entanglement measures $S_x=-\sum_k(\nu_k^{(x)}/\mathcal{N})\ln(\nu_k^{(x)}/\mathcal{N})$ for $x=R,z,Q$. Here $\nu_k^{(x)}$ are the natural populations of mode $x$.  Figure~\ref{fig:entropy}(b) shows these entropies for $g=0.12$~a.u. At $t=0$, only the parametric $R$--$z$ correlation of the Franck--Condon state $\chi_1(R)\,\xi_2(z;R)\,|0\rangle$ is present ($S_R=S_z\ll1$, $S_Q=0$). The photon becomes entangled with the molecule within the first femtosecond, and $S_Q$ stays between $0.74$ and $0.83$ during 3--12~fs, close to the value $-p_{\xi_1}\ln p_{\xi_1}-(1-p_{\xi_1})\ln(1-p_{\xi_1})=0.69$ of the entangled state $\sqrt{1-p_{\xi_1}}\,|0,\xi_2\rangle+\sqrt{p_{\xi_1}}\,|1,\xi_1\rangle$ (nuclear factors omitted), with $p_{\xi_1}=0.44$ from Table~\ref{tab:dynamics}.

Computing $S_R$ and $S_z$ in the multistate representation would require the coherences between electronic states; for the photon the bounds of Eq.~\eqref{eq:entropy_bounds} are available (Sec.~\ref{sec:observables}). The $S_Q$ of \textsf{enpd} lies at all times between the two bounds of Eq.~\eqref{eq:entropy_bounds} evaluated with \textsf{npd--fDSE--MLS} (Fig.~\ref{fig:entropy}(b)). For $1\le t\le22$~fs, the entanglement generated within the electronic states, $\bar{S}$, accounts for only 9--43\% of $S_Q$ (filtered curves); in the multistate model, the remainder, $S_Q-\bar{S}$, measures how different the photonic states of the electronic channels are (Sec.~\ref{sec:observables}). The light--matter entanglement is thus dominated by the correlation between the photon number and the electronic state{, which is created by the radiative transfer within the first femtosecond ($S_Q-\bar{S}\approx0.57$ at 1~fs), is extended to higher states by sequential photon exchange, and} is not contained in the conditional entropies $S_i$. While $S_R$ and $S_z$ grow monotonically ($S_R=1.15$ and $S_z=1.26$ at 20~fs), $S_Q$ rises to $0.83$ at 6~fs, decreases to $0.74$ at 12~fs (when $S_Q-\bar{S}$ drops and $\bar{S}$ grows) and increases again to $0.96$ at 20~fs, as the $R$-dependent displacement of the polarization field and the population of higher states correlate the photon with the nuclear motion. The electron is the most entangled mode, because it is correlated both with the nuclei and with the photon. For $t\gtrsim22$~fs (dotted line in Fig.~\ref{fig:entropy}(b)), the state conditioned on survival is affected by the absorption at the CAP, and the late changes of the entropies are not physical.

%% file: section_discussion.tex
\section{Discussion}
\label{sec:discussion}

\subsection{Relation to previous work}
\label{sec:disc_previous}

The suppression of a photochemical reaction by a cavity resonant with an electronic transition was predicted by Galego \textit{et al.}~\cite{Galego2016}, who showed that the polaritonic surfaces can decouple the excited-state population from the reactive pathway. The mechanism found here is different in its nature and in its control parameter. The packet is not steered by a modified excited-state surface; instead, the cavity provides a radiative channel, $|0,\xi_2\rangle\to|1,\xi_1\rangle$, that is active only in the narrow resonance region around $R_{\text{eq}}$, and the stabilization yield is set by the ratio between the vacuum Rabi splitting and the rate at which the packet leaves that region. This is apparent in the crossover from diabatic to adiabatic passage at $g\approx0.02$--$0.03$~a.u. (Fig.~\ref{fig:fock}(a)), i.e., at a normalized coupling $\eta_{\text{R}}\approx0.03$--$0.04$ (Table~\ref{tab:couplings}), and a saturation of the yield, set by the weight of the {upper} polaritonic branch. The closest antecedent is the cavity-induced photostabilization of the dissociating MgH$^+$ ion by Davidsson and Kowalewski~\cite{Davidsson2020}, who propagated the nuclear packet with a Lindblad equation for a lossy cavity and found stabilization both for long photon lifetimes and for lifetimes below 1~fs. The present work addresses the closed-cavity limit of such a process and focuses on the role of the PDMs, the DSE, and the electronic states beyond the two-level manifold. The modification of dissociation in an exactly solved molecule~\cite{Sidler2020} and the cavity-induced changes of molecular structure~\cite{Galego2015,Flick2017cBO} share with the present work the conclusion that the nuclear degrees of freedom must be treated dynamically; the present results add the role of the PDMs and of electronic states beyond the TLS, which together determine both the photonic observables and the fraction of the population that escapes the two-level manifold.

Double-well structures along the photon displacement coordinate have been reported in cavity Born--Oppenheimer and exact-factorization studies~\cite{Flick2017cBO,Abedi2018}. In the polaritonic-surface partition adopted here, which is the natural one for a cavity resonant with an electronic transition~\cite{ruggenthaler_understanding_2023}, the relevant objects are the polaritonic PECs of Fig.~\ref{fig:pecs}, and the coupled nucleus--photon dynamics is solved without invoking the polaritonic partition. The photonic densities of Figs.~\ref{fig:rho_g001} and \ref{fig:rho_g012} are centered at state-dependent displacements that follow the PDM of each electronic state, as expected from Eq.~\eqref{eq:completed_square}, and their nodal structure reflects the photon number (e.g., the one-photon node in $\xi_1$). A  symmetric double well in $Q$ requires a symmetric model with a strong TDM and no PDMs, and is the analog, for two electronic states coupled through a displacement coordinate, of an avoided crossing; in a heteronuclear molecule it is replaced by the state-dependent displacements found here.

\subsection{Role of the dipole self-energy and experimental realizability}
\label{sec:disc_feasibility}

Within the single-mode PF model, the DSE is not optional: its coefficient $g^2/\hbar\omega_0=\lambda^2/2$ is fixed by the PZW transformation, and the Proposition of Sec.~S1 of the Supplemental Material~\cite{SM} shows that this value is also necessary for the spectrum of the truncated Hamiltonian to be independent of the dipole origin. Models that omit or rescale the DSE, common in the cavity-QED literature, therefore break this invariance and produce an $R$-dependent residual that, along the dissociation coordinate, either removes the ground state or adds a spurious confinement. The quantitative effect of the DSE on the dynamics is moderate in the coupling range studied ($E_{\text{q}}\le2.2$~eV, and $|E^{\text{fDSE}}_\alpha-E^{\text{tDSE}}_\alpha|\le10^{-3}$~a.u.\ for the populated branches), but it is not small compared with the vibrational energy scale, and its omission would bias precisely the large-$R$ region where the PDMs are largest. Fern\'andez de la Pradilla \textit{et al.}~\cite{Pradilla2025} have argued that, in nanophotonic structures, USC originates from the quasistatic Coulomb interaction between the emitter and the charges of the structure rather than from transverse photons, and that the polarization self-energy then need not be included. The present results do not contradict this argument: they show that, once a single effective transverse mode is adopted, as in most of the molecular cQED literature, the DSE is required for internal consistency{, in line with ab initio treatments of general electromagnetic environments, where the DSE is needed for gauge invariance and for the stability of the coupled system~\cite{Svendsen2024}. Moreover, in the quasistatic limit, where the PZW transformation leaves the emitter--field interaction unchanged~\cite{Feist2021}, the stabilizing role of the quadratic term is taken over by the self-polarization of the plasmonic structure~\cite{Kuisma2022}.} A treatment of the quasistatic near field of a specific nanostructure would replace the single-mode coupling by a different Hamiltonian, and is outside the scope of this work. Exact simulations of molecular ensembles that treat electrons, nuclei, and photons on an equal footing have recently addressed the DSE beyond the Born--Oppenheimer approximation~\cite{Krupp2026}.

Table~\ref{tab:couplings} gives the mode volumes that correspond to the couplings studied, between $6$ and $0.04\ \text{nm}^3$. Since $g\propto(\omega_0/V)^{1/2}$, these volumes refer to $\hbar\omega_0=17.6$~eV, and the comparison with experiments is made through $\eta_{\text{R}}$, which, together with the slope of the detuning at the resonance and the velocity of the packet, controls the mechanism. Single methylene-blue molecules in plasmonic nanogaps with $V<40\ \text{nm}^3$ show vacuum Rabi splittings of 90~meV at $\hbar\omega\approx1.9$~eV~\cite{Chikkaraddy2016}, i.e., $\eta_{\text{R}}\approx0.02$, comparable to the smallest couplings studied here ($g=0.014$--$0.02$~a.u.){, for which the stabilization yield is already $22$--$33\%$ (Table~\ref{tab:dynamics}). The mechanism therefore does not require the ultrastrong regime: a sizable fraction of the dissociating population is trapped at couplings that have been reached for single molecules.} Picocavities formed by atomic features in such gaps confine the field to volumes well below $1\ \text{nm}^3$~\cite{Benz2016}; for the same transition, the scaling $\eta_{\text{R}}\propto V^{-1/2}$ gives $\eta_{\text{R}}\sim0.15$, at the upper end of the range studied, although electronic strong coupling has not been demonstrated in these transient structures. First-principles calculations nevertheless predict single-molecule USC, with $g/\omega_{\text{ex}}\approx0.11$--$0.13$ ($\omega_{\text{ex}}$ the molecular transition frequency), in plasmonic nanoparticle dimers with mode volumes of $1$--$10\ \text{nm}^3$~\cite{Kuisma2022}, without resorting to the atomic-scale features of picocavities~\cite{Urbieta2018,Baumberg2019}; this corresponds to $g\approx0.08$--$0.10$~a.u.\ in Table~\ref{tab:couplings}. Diffraction-limited Fabry--P\'erot or fiber cavities, with $V\gtrsim\lambda_{\text{c}}^3$, where $\lambda_{\text{c}}$ is the wavelength of the mode, give single-emitter couplings several orders of magnitude smaller. These comparisons fix the coupling scale; in nanogaps and picocavities the coupling is mediated mainly by the quasistatic near field, whose description beyond a single transverse mode was discussed above~\cite{Pradilla2025}. NeHe$^+$ plays here the role of a model system: its transition lies in the extreme ultraviolet, above the range of plasmonic cavities, which with aluminum or magnesium extends to $\sim7$--$8$~eV~\cite{Kuisma2022,TorresSanchez2021}, but the mechanism relies only on generic features, a dissociative charge-transfer state resonant with the cavity in the Franck--Condon region, a large transition dipole at the resonance, and PDMs that grow along the dissociation coordinate. These features are not specific to NeHe$^+$ but generic for charge-transfer transitions of heteronuclear molecular ions, some of which lie at lower energies; for such systems, the present results indicate the behavior to be expected when comparable values of $\eta_{\text{R}}$ and $\gamma_{\text{LZ}}$ are reached.

A second ingredient of real nanocavities is their loss. Plasmonic cavities have quality factors $Q_{\text{c}}$ of order 10~\cite{Baumberg2019,TorresSanchez2021}, i.e., a photon decay rate $\kappa=\omega_0/Q_{\text{c}}$ that is only one order of magnitude below the mode frequency. The vacuum Rabi splitting exceeds the cavity linewidth, $2g|\mu^{\text{BOA}}_{12}(R_{\text{eq}})|>\hbar\kappa$, only for $\eta_{\text{R}}>1/(2Q_{\text{c}})\approx0.05$, i.e., for $g\gtrsim0.04$~a.u.\ in Table~\ref{tab:couplings}; the smaller couplings would therefore be in the weak-coupling regime of such a cavity. Moreover, the photon lifetime $Q_{\text{c}}/\omega_0$ spans only $Q_{\text{c}}/2\pi\approx1.6$ optical cycles, i.e., a few femtoseconds for visible or near-infrared transitions, comparable to the $\sim3$~fs over which the radiative stabilization takes place here. Losses, however, do not act against the mechanism. Since the loss of the photon leaves the molecule in the bound state $|0,\xi_1\rangle${, near the equilibrium distance of $\xi_1$ (unlike H$_2$ near an aluminum nanoparticle, where the plasmon decay triggers dissociation on the ground state~\cite{TorresSanchez2021})}, it does not undo the stabilization; it removes instead the coherent back-transfer $|1,\xi_1\rangle\to|0,\xi_2\rangle$ and turns the coherent passage described in Sec.~\ref{sec:res_densities} into an irreversible emission{. This is the behavior found for MgH$^+$, whose stabilization is reduced for photon lifetimes of tens of femtoseconds but increases again below 1~fs~\cite{Davidsson2020}, and for the photoprotection of uracil, which is most efficient at a finite photon decay rate~\cite{Felicetti2020}.

The yield may nevertheless differ from the one computed here; its assessment requires an open-system treatment, e.g., a Lindblad dissipator acting on the cavity mode~\cite{Davidsson2020}, constructed in the dressed light--matter basis at the largest couplings~\cite{Beaudoin2011,Settineri2018}. In such a treatment, the escaping photon would carry the correlation between photon number and molecular fate found in Sec.~\ref{sec:res_entanglement}, so that losses are expected to transfer the light--matter entanglement to the output field rather than to erase it.}
Taken together, these estimates place the couplings studied between those demonstrated for single molecules in plasmonic nanogaps and those expected in picocavities. The present single-mode PF calculations identify the stabilization mechanism and its dimensionless control parameters; combining them with cavity losses and with a realistic description of the near field~\cite{Feist2021,Svendsen2024} is the natural next step toward a quantitative prediction for a specific nanostructure.

\subsection{Limitations}
\label{sec:disc_limitations}
Besides the absence of losses, discussed above, the ten-state model rests on four approximations. The first two are validated by the calculations presented here; the last two leave open questions of a technical nature that do not affect the main conclusions. (i) The bare electronic nonadiabatic couplings $d_{ij}$ are neglected in \textsf{npd--fDSE--MLS}; the cavity-induced couplings are treated exactly. The \textsf{enpd} propagation, which includes them, shows that this approximation, together with the truncation to ten states, leaves the populations and the mean free-photon number nearly unchanged at the largest coupling; $\mathcal{G}^{(2)}_{\text{PZW}}(0)$ also agrees in order of magnitude once the numerical tail of the ten-state distribution is removed (Sec.~\ref{sec:res_statistics}). (ii) The molecule is described by a one-dimensional single-active-electron model{, coupled to the cavity after the dimensional reduction. Reducing the dimension before or after the coupling gives the same result in the complete-basis limit, and the pQED--sepd comparison confirms this agreement for the polaritonic PECs.} (iii) The electronic space is truncated to $N_{\text{PECs}}=10$ states; the convergence of the DSE block is controlled by $\Delta\mathbb{M}^2$,{ and the last convergence step of the polaritonic PECs is of the same order as the fDSE--tDSE difference (Appendix~\ref{app:trk}). At this level of truncation both forms of the DSE are therefore equally accurate, and which of them converges faster to the complete limit is left for larger electronic bases.} (iv) The free-photon distribution of the ten-state model has a small numerical tail that extends to the edge of the photon-number basis. It hardly changes $\langle\hat{n}_{\text{PZW}}\rangle$ but dominates $\mathcal{G}^{(2)}_{\text{PZW}}(0)$ at the largest couplings, and we remove it with a geometric continuation of the physical part of the distribution (Sec.~\ref{sec:res_statistics}). The comparison with \textsf{enpd} shows that the tail is an artifact of the ten-state calculation (Appendix~\ref{app:photon}).

%% file: section_conclusions.tex
\section{Conclusions}
\label{sec:conclusions}

The coupled nucleus--photon dynamics of a one-dimensional NeHe$^+$ model in a single cavity mode resonant with its first electronic excitation has been studied. A ten-state Pauli--Fierz Hamiltonian in the length gauge was developed that retains the full DSE, permanent and transition dipoles.

 It was shown that for the truncated multistate Hamiltonian, the DSE coefficient $g^2/\hbar\omega_0$ remains necessary and sufficient for the invariance of the spectrum under a change of dipole origin, in both its truncated (tDSE) and full (fDSE) forms. The difference between the two forms is positive semidefinite, bounded by the missing TRK oscillator strength, and concentrated in the last retained state; for the populated polaritonic branches during the dynamics it only amounts to at most $10^{-3}$~a.u.

The polaritonic PECs in the photon-number basis agree with a simultaneous electron--photon diagonalization free of electronic truncation to within $1.3\times10^{-3}$~a.u.\ for the three lowest branches at the largest coupling, compared with $4\times10^{-4}$~a.u.\ at the weakest one, where the difference reflects only the electronic discretization. On the other hand, the TLS changes up to $1.3\times10^{-2}$~a.u.\ at the largest coupling, mainly at the near-resonant third state.

The  presence of the cavity opens a radiative stabilization channel at the resonant avoided crossing between $|0,\xi_2\rangle$ and $|1,\xi_1\rangle$, which quenches the dissociation of the excited molecule. The trapped population grows from $14\%$ to $49\%$ between $g=0.01$ and $0.06$~a.u., following the crossover from diabatic to adiabatic passage, and saturates at the weight of the {upper} polaritonic branch. For $g\gtrsim0.08$~a.u., more than $7\%$ of the population leaves the two-level manifold through sequential photon exchange, suggesting that it is no longer a good approximation.

A thorough analysis of the photonic distributions was performed. The length-gauge coherence $\langle\hat{a}\rangle$ is dominated by the static polarization field generated by the PDMs. The free-photon coherence $\langle\hat{a}_{\text{PZW}}\rangle$ is an order of magnitude smaller, the mean free-photon number saturates at $\simeq0.5$, close to the stabilization yield. The free-photon statistics is sub-Poissonian, dominated by the single photon emitted in the stabilization. In the length gauge, the multiphoton content of the field is produced by the polarization fields of the PDMs. The nucleus--photon entanglement within the electronic states is small in the bound channel and dominated by the dissociative one, although both grow monotonically with $g$, by two orders of magnitude between $g=0.01$ and $0.12$~a.u.

A full electron--nucleus--photon propagation at the largest coupling, free of electronic truncation and including the nonadiabatic couplings, the continuum, and the mass polarization, reproduces the surviving probability and the mean free-photon number of the ten-state model. This full dynamics  further reveals a weak electronic ionization channel. It also {gives access to the} light--matter entanglement{,} which {is} dominated by the correlation between photon number and electronic state created by the radiative stabilization. This entanglement is generated within the first femtosecond by the radiative transfer, and its magnitude is set by the stabilization yield: the photon entropy, $0.74$--$0.83$ during 3--12~fs, is close to the binary entropy $0.69$ of $p_{\xi_1}=0.44$, which approaches its maximum, $\ln2$, as the yield saturates near $1/2$. In the closed cavity this entanglement is retained, so that the photon number carries the information on whether the molecule was trapped or dissociates. The free-photon second-order coherence{ of this propagation, $\mathcal{G}^{(2)}_{\text{PZW}}(0)\simeq0.07$ on average over 5--18~fs, is} of the same order as the ten-state value once the numerical tail of its photon-number distribution is removed ($0.09$ {on average}): the field of the cavity is close to a mixture of zero and one photon.

The stabilization does not require the ultrastrong regime: at the couplings demonstrated for single molecules in plasmonic nanogaps ($\eta_{\text{R}}\approx0.02$), the yield is already 22--33\%. A natural extension of this work is an open-system treatment of cavity losses and of the output field, which are expected to render the stabilization irreversible rather than to suppress it~\cite{Davidsson2020}. Also, a description of the quasistatic near field of specific nanostructures (beyond the single-mode model~\cite{Feist2021,Svendsen2024}) could shed some light on the strength of the coupling. Exploratory analysis and the identification of heteronuclear molecular ions with charge-transfer transitions at lower energies and large transition dipoles, for which the dimensionless parameters {$\eta_{\text{R}}$ and $\gamma_{\text{LZ}}$}, that control the radiative stabilization, can be reached in plasmonic nanocavities~\cite{Chikkaraddy2016,Kuisma2022}.

%% file: section_acknowledgments.tex
\begin{acknowledgments}
Partial financial support from CONICET (PIP-KE311220210100787CO) and SECYT-UNC (PIDTA 2026 - 33820250100211CB) is gratefully acknowledged. M.M. acknowledges financial support from CONICET through a doctoral fellowship. This work used computational resources from CCAD -- Universidad Nacional de C\'ordoba (\url{https://ccad.unc.edu.ar/}), in particular the Mulatona and Serafín clusters, which are part of SNCAD -- MinCyT, Rep\'ublica Argentina.

\textit{Author contributions.} M.M.\ developed the FEMTISE software and performed all the electronic-structure, polaritonic, and MCTDH calculations. M.M.\ and F.M.P.\ contributed equally to the conceptualization, the formal analysis, and the writing of the manuscript.
\end{acknowledgments}

%% file: appendix_hamiltonians.tex
\section{Hamiltonians of the model hierarchy}
\label{app:hamiltonians}

This appendix collects the Hamiltonians of Fig.~\ref{fig:models} that are not written explicitly in the main text. The reference \textsf{enpd} propagates Eq.~\eqref{eq:PF_ham} in $(R,z,Q)$ with the mass-polarization terms of Sec.~\ref{sec:methods_full},
\begin{equation}\label{eq:H_enpd}
  \begin{split}
  \hat{H}_{\text{enpd}}={}&\hat{T}_R+\frac{\hbar^2s}{\mathcal{M}_{\text{red}}}\partial_R\partial_z-\frac{\hbar^2s^2}{2\mathcal{M}_{\text{red}}}\partial_z^2+\hat{H}_{\text{el}}(z;R)+\frac{\hat{P}^2}{2}\\
  &+\frac{\omega_0^2\hat{Q}^2}{2}+g\sqrt{\frac{2\omega_0}{\hbar}}\,\hat{Q}\,\hat{\mu}_\epsilon+\frac{g^2}{\hbar\omega_0}\hat{\mu}_\epsilon^{\,2},
  \end{split}
\end{equation}
with $\hat{\mu}_\epsilon=\mathcal{Q}_{\text{mid}}R-|e|\hat{z}$. In \textsf{sepd} and \textsf{pQED}, the conditional problem at fixed $R$ (Eq.~\eqref{eq:V2D}, or Eqs.~\eqref{eq:An} and \eqref{eq:Cn} with $\mathbb{S}=\mathbb{D}$ (\textsf{fDSE}) or $\mathbb{M}^2$ (\textsf{tDSE}) and $N_{\text{PECs}}=10$ (\textsf{MLS}) or 2 (\textsf{TLS})) gives the polaritonic PECs, and the nuclear dynamics on each of them, neglecting the nonadiabatic couplings between polaritonic surfaces, is governed by
\begin{equation}\label{eq:H_pec}
  \hat{H}_\alpha=\hat{T}_R+E_\alpha(R).
\end{equation}
The two-level models without PDMs follow from Eq.~\eqref{eq:PF_MLS_ham} with $N_{\text{PECs}}=2$ and $\mu^{\text{BOA}}_{ii}\equiv0$, for which $\mathbb{M}^2=(\mu^{\text{BOA}}_{12})^2\hat{\sigma}_0$ (Eq.~\eqref{eq:tls_M2}):
\begin{equation}\label{eq:H_vRabi}
  \begin{split}
  \hat{H}_{\text{vR}}={}&\hat{T}_R+\bar{\epsilon}(R)\,\hat{\sigma}_0+\frac{\Delta\epsilon_{12}(R)}{2}\hat{\sigma}_3+\frac{\hat{P}^2}{2}+\frac{\omega_0^2\hat{Q}^2}{2}\\
  &+g\sqrt{\frac{2\omega_0}{\hbar}}\,\hat{Q}\,\mu^{\text{BOA}}_{12}(R)\,\hat{\sigma}_1+\eta_{\text{DSE}}\big[\mu^{\text{BOA}}_{12}(R)\big]^2\hat{\sigma}_0,
  \end{split}
\end{equation}
with $\bar{\epsilon}=(\epsilon_1+\epsilon_2)/2$, and $\eta_{\text{DSE}}=\eta_{\text{req}}\equiv g^2/\hbar\omega_0$ for \textsf{npd--tDSE--TLS--noPDM} or $\eta_{\text{DSE}}=0$ for \textsf{npd--noDSE--TLS--noPDM}. For $\eta_{\text{DSE}}=0$ and fixed $R$, Eq.~\eqref{eq:H_vRabi} reduces, up to a constant, to the quantum Rabi model of Eq.~\eqref{eq:PF_Rabi_h}, and the rotating-wave approximation gives the Jaynes--Cummings model,
\begin{equation}\label{eq:H_JC}
  \hat{H}_{\text{JC}}=\frac{\Delta\epsilon_{12}}{2}\hat{\sigma}_3+\hbar\omega_0\hat{a}^{\dagger}\hat{a}+g\mu^{\text{BOA}}_{12}\big(\hat{a}\,\hat{\sigma}_++\hat{a}^{\dagger}\hat{\sigma}_-\big),
\end{equation}
with $\hat{\sigma}_+=|\xi_2\rangle\langle\xi_1|$ and $\hat{\sigma}_-=\hat{\sigma}_+^\dagger$.

%% file: appendix_trk.tex
\section{Thomas--Reiche--Kuhn bound on the dipole self-energy gap}
\label{app:trk}

For the one-electron Hamiltonian $\hat{H}_{\text{el}}=\hat{p}^2/2\mathcal{M}_e+V_{\text{el}}(\hat{z};R)$, the double commutator $[\hat{z},[\hat{H}_{\text{el}},\hat{z}]]=\hbar^2/\mathcal{M}_e$ evaluated in the eigenstate $\xi_i$ gives the TRK sum rule~\cite{Thomas1925,ReicheThomas1925,Kuhn1925}
\begin{eqnarray}\label{eq:trk}
  \sum_j(\epsilon_j-\epsilon_i)|z_{ij}|^2 &=& \frac{\hbar^2}{2\mathcal{M}_e}, \nonumber\\
  \sum_jf_{ij} &=& 1, \nonumber\\
  f_{ij} &=& \frac{2\mathcal{M}_e}{\hbar^2}(\epsilon_j-\epsilon_i)|z_{ij}|^2 .
\end{eqnarray}
Splitting the sum into retained and excluded states and bounding $\epsilon_j-\epsilon_i\ge\epsilon_{N_{\text{PECs}}+1}-\epsilon_i>0$ for $j>N_{\text{PECs}}$ yields Eq.~\eqref{eq:Delta_trk_bound} for $(\Delta\mathbb{M}^2)_{ii}=\sum_{j>N_{\text{PECs}}}|z_{ij}|^2$. The same distribution of weights $|z_{ij}|^2$ determines the missing oscillator strength, the missing static polarizability, and $(\Delta\mathbb{M}^2)_{ii}$: they are different moments of the excluded spectral distribution and vanish together as $N_{\text{PECs}}\to\infty$. Because $f_{ij}=-f_{ji}$, $\sum_{i\le N_{\text{PECs}}}\mathcal{F}_i^{(N_{\text{PECs}})}=0$, so $\sum_i[1-\mathcal{F}_i^{(N_{\text{PECs}})}]=N_{\text{PECs}}$: when the low states nearly saturate the sum rule, the deficit accumulates in $\xi_{N_{\text{PECs}}}$, and since the gap $\epsilon_{N_{\text{PECs}}+1}-\epsilon_{N_{\text{PECs}}}$ is small in the discretized continuum, Eq.~\eqref{eq:Delta_trk_bound} leaves $(\Delta\mathbb{M}^2)_{N_{\text{PECs}} N_{\text{PECs}}}$ essentially unbounded. Table~\ref{tab:dse_gap} verifies these statements at $R_{\text{eq}}$. The bound, evaluated with $\epsilon_{N_{\text{PECs}}}$ instead of $\epsilon_{N_{\text{PECs}}+1}$ (which weakens it but keeps it valid for $i<N_{\text{PECs}}$), holds at all 1981 grid points.

\begin{table}[htbp]
\caption{\label{tab:dse_gap}DSE gap for $N_{\text{PECs}}=10$ at $R_{\text{eq}}\simeq1.67$: $\langle z^2\rangle_i=\langle\xi_i|\hat{z}^2|\xi_i\rangle$ (with $z$ measured from the internuclear midpoint), partial oscillator-strength sum $\mathcal{F}_i\equiv\mathcal{F}_i^{(10)}$, $(\Delta\mathbb{M}^2)_{ii}$ of Eq.~\eqref{eq:mu2_closure}, the bound of Eq.~\eqref{eq:Delta_trk_bound}, and the first-order energy shift $\delta E_i=\eta_{\text{req}} e^2(\Delta\mathbb{M}^2)_{ii}$ for $g=0.12$~a.u.\ (atomic units).}
\begin{ruledtabular}
\begin{tabular}{rrrrrr}
$i$ & $\langle z^2\rangle_i$ & $\mathcal{F}_i$ & $(\Delta\mathbb{M}^2)_{ii}$ & bound & $\delta E_i$ \\
\hline
1  & 0.778 & 0.991  & $1.3\times10^{-3}$ & $2.5\times10^{-3}$ & $2.8\times10^{-5}$ \\
2  & 2.29  & 0.886  & $3.7\times10^{-2}$ & $4.8\times10^{-2}$ & $8.1\times10^{-4}$ \\
3  & 8.45  & 0.973  & $1.5\times10^{-2}$ & $2.3\times10^{-2}$ & $3.2\times10^{-4}$ \\
4  & 25.4  & 0.879  & $9.5\times10^{-2}$ & 0.15 & $2.1\times10^{-3}$ \\
5  & 73.1  & 0.991  & $4.7\times10^{-3}$ & $1.4\times10^{-2}$ & $1.0\times10^{-4}$ \\
6  & 92.8  & 0.869  & 0.14 & 0.22 & $3.1\times10^{-3}$ \\
7  & 87.7  & 0.893  & 0.10 & 0.23 & $2.2\times10^{-3}$ \\
8  & 87.8  & 0.821  & 0.15 & 0.51 & $3.2\times10^{-3}$ \\
9  & 85.0  & 0.762  & 0.26 & 1.4 & $5.6\times10^{-3}$ \\
10 & 84.7  & $-8.06$ & 37.9 & --- & 0.82 \\
\end{tabular}
\end{ruledtabular}
\end{table}

The off-diagonal elements $(\Delta\mathbb{M}^2)_{ik}$ couple electronic states within the same photon-number sector, because the DSE is diagonal in $n$; their effect on state $i$ is of second order, $\sim|\eta_{\text{req}} e^2(\Delta\mathbb{M}^2)_{ik}|^2/|\epsilon_i-\epsilon_k|$, which is $\sim4\times10^{-9}$~a.u.\ for the pair $(1,2)$ at $R_{\text{eq}}$ (Fig.~\ref{fig:dipoles}(f)). To first order, the resonant crossing is displaced from $R_{\text{eq}}$ by $\delta R\approx-\eta_{\text{req}}((\Delta\mathbb{M}^2)_{22}-(\Delta\mathbb{M}^2)_{11})/\partial_R\Delta\epsilon_{12}(R_{\text{eq}})=1.7\times10^{-3}$~a.u.\ for $g=0.12$~a.u.\ (Table~\ref{tab:dse_gap}), below the resolution $\Delta R=0.01$~a.u.\ with which it was located (Sec.~\ref{sec:initial_state}).

The convergence of the polaritonic PECs with $N_{\text{PECs}}$ was assessed by matching the branches $\alpha\le3$ of $N_{\text{PECs}}<10$ to those of $N_{\text{PECs}}=10$ with fDSE by maximum eigenvector overlap (overlap $\ge0.97$ for $N_{\text{PECs}}\ge4$) on $R\in[0.88,10]$~a.u., the region explored by the packet. With the fDSE the convergence is monotonic, the change in the last step ($N_{\text{PECs}}=9\to10$) grows with $g$, and the tDSE converges to values that differ from the fDSE ones by an amount of the same order. For small $N_{\text{PECs}}$ the tDSE is closer to the $N_{\text{PECs}}=10$ reference than the fDSE with the same $N_{\text{PECs}}$, because it omits both the linear coupling to the excluded states and their DSE, whose second-order contributions, $-g^2|\mu^{\text{BOA}}_{im}|^2/(\epsilon_m-\epsilon_i+\hbar\omega_0)$ and $+g^2|\mu^{\text{BOA}}_{im}|^2/\hbar\omega_0$, partially cancel, whereas the fDSE includes only the second one.

%% file: appendix_photon.tex
\section{Length-gauge and free-photon observables}
\label{app:photon}

By Eq.~\eqref{eq:physical_photon}, the free-photon annihilation operator differs from the length-gauge one by a matter operator, $\hat{a}_{\text{PZW}}=\hat{a}+(g/\hbar\omega_0)\hat{\mu}_\epsilon$. The sign of $g$ is fixed by the orientation of $\vec{\epsilon}_0$ along the molecular axis and is therefore a convention: reversing $\vec{\epsilon}_0$ is equivalent to reversing the orientation of the molecule with respect to the mode. Since $\hat{\Pi}_{\text{ph}}$ (Eq.~\eqref{eq:parity}) maps $\hat{H}(g)$ onto $\hat{H}(-g)$ and leaves the initial length-gauge vacuum invariant, a change of convention maps $\hat{a}\to-\hat{a}$ and, because $\hat{\mu}_\epsilon$ commutes with $\hat{a}$, $\hat{a}_{\text{PZW}}\to-\hat{a}_{\text{PZW}}$. Hence $\langle\hat{n}\rangle$, $\langle\hat{n}_{\text{PZW}}\rangle$, $\mathcal{G}^{(2)}_{\text{PZW}}(0)$, and the entropies are independent of it, whereas $\langle\hat{a}\rangle$, $\langle\hat{Q}\rangle$, and the intracavity field change sign: only the sign of $\langle\hat{a}\rangle$ relative to $g\langle\hat{\mu}_\epsilon\rangle$ is meaningful. If the molecular dipole varies slowly on the scale of the photon period $2\pi/\omega_0$, the field adiabatically follows it, and the field part of Eq.~\eqref{eq:PF_ham}, $\hbar\omega_0\hat{a}^{\dagger}_{\text{PZW}}\hat{a}_{\text{PZW}}$, is minimized by $\langle\hat{a}_{\text{PZW}}\rangle=0$, i.e., by Eq.~\eqref{eq:static_estimate}.

Table~\ref{tab:static} compares Eq.~\eqref{eq:static_estimate}, evaluated with the electronic populations $p_{\xi_i}(t)/\mathcal{N}(t)$ and the mean distances $\langle R\rangle_i(t)$ from the MCTDH propagation\ and the PDMs $\mu^{\text{BOA}}_{ii}(R)$ of Fig.~\ref{fig:dipoles}(a) at these distances, with the computed $\text{Re}\,\langle\hat{a}\rangle$. The instantaneous values oscillate at the photon period around the static estimate; averaged over one photon period, the relative deviation for $2\le t\le30$~fs is at most $5\%$ for $g=0.01$~a.u., $10\%$ for $g=0.06$~a.u., and $17\%$ for $g=0.08$~a.u., with medians between $1\%$ and $5\%$ for all couplings. The static polarization field therefore accounts for most of the length-gauge coherence.

\begin{table}[htbp]
\caption{\label{tab:static}Real part of the length-gauge coherence, $\text{Re}\,\langle\hat{a}\rangle$ (computed), and the static estimate of Eq.~\eqref{eq:static_estimate}, at selected times $t$ (fs).}
\begin{ruledtabular}
\begin{tabular}{cccccc}
$g$ (a.u.) & & $t=5$ & $t=10$ & $t=20$ & $t=28$ \\
\hline
0.01 & computed & $-0.043$ & $-0.097$ & $-0.141$ & $-0.057$ \\
 & Eq.~\eqref{eq:static_estimate} & $-0.035$ & $-0.082$ & $-0.156$ & $-0.075$ \\
0.04 & computed & $-0.131$ & $-0.191$ & $-0.371$ & $-0.160$ \\
 & Eq.~\eqref{eq:static_estimate} & $-0.108$ & $-0.218$ & $-0.390$ & $-0.205$ \\
0.12 & computed & $-0.336$ & $-0.546$ & $-0.993$ & $-1.025$ \\
 & Eq.~\eqref{eq:static_estimate} & $-0.290$ & $-0.538$ & $-1.001$ & $-1.087$ \\
\end{tabular}
\end{ruledtabular}
\end{table}

The decrease of $|\text{Re}\,\langle\hat{a}\rangle|$ at late times for the smaller couplings reflects the absorption of the dissociating fragments, which carry the largest PDMs, and the corresponding increase of the relative weight of the bound fraction in $\xi_1$, whose PDM is small. The same decomposition applies to the photon number: $\langle\hat{n}\rangle$ contains, besides $\langle\hat{n}_{\text{PZW}}\rangle$, the cross term and the term $(g/\hbar\omega_0)^2\langle\hat{\mu}_\epsilon^2\rangle$ associated with the polarization, which grows as the fragments separate.

\paragraph{Free-photon populations.} The distribution $p_n^{\text{phys}}$ was obtained by transforming the MCTDH wave function to the PZW representation, $\Psi\to\hat{U}_{\text{PZW}}^\dagger\Psi$ (Eq.~\eqref{eq:physical_photon}), on the full $(R,Q)$ grid. At each grid point, the electronic components are expressed in the eigenbasis $\{|k\rangle\}$ of the dipole matrix $\mathbb{M}(R)$ (Eq.~\eqref{eq:dipole_matrix}), with eigenvalues $m_k(R)$; the photonic component of each $|k\rangle$ is displaced by $\beta m_k(R)$, with $\beta=g/\hbar\omega_0$, and projected onto the harmonic-oscillator basis of $Q$, which gives $p_n^{\text{phys}}$. As a check, $\langle\hat{a}_{\text{PZW}}\rangle$ and $\langle\hat{n}_{\text{PZW}}\rangle$ obtained in this way agree to $10^{-11}$ with those computed from the moments of the untransformed state through Eq.~\eqref{eq:physical_photon}, $\langle\hat{a}_{\text{PZW}}\rangle=\langle\hat{a}\rangle+\beta\langle\mathbb{M}\rangle$ and $\langle\hat{n}_{\text{PZW}}\rangle=\langle\hat{n}\rangle+\beta\langle(\hat{a}+\hat{a}^{\dagger})\mathbb{M}\rangle+\beta^2\langle\mathbb{M}^2\rangle$.

\paragraph{Powers of the dipole operator.} In the ten-state model, the powers of $\hat{\mu}_\epsilon$ that appear in $\hat{a}_{\text{PZW}}^{\dagger\,2}\hat{a}_{\text{PZW}}^{2}$ can be represented either by the matrices of the powers, $(\mathbb{D}_\ell)_{ij}=\langle\xi_i|\hat{\mu}_\epsilon^{\,\ell}|\xi_j\rangle$, with $\mathbb{D}_2=\mathbb{D}$ the fDSE matrix, or by the powers $\mathbb{M}^\ell$ of the dipole matrix. With $\hat{\mathcal{P}}$ the projector on the retained states, $\hat{\mathcal{P}}_{\perp}=1-\hat{\mathcal{P}}$, $B=\hat{\mathcal{P}}\hat{\mu}_\epsilon\hat{\mathcal{P}}_{\perp}$, and $C=\hat{\mathcal{P}}_{\perp}\hat{\mu}_\epsilon\hat{\mathcal{P}}_{\perp}$,
\begin{eqnarray}\label{eq:dipole_powers}
 \mathbb{D}_2&=&\mathbb{M}^2+BB^{\mathsf{T}},\\
 \mathbb{D}_3&=&\mathbb{M}^3+\mathbb{M}BB^{\mathsf{T}}+BB^{\mathsf{T}}\mathbb{M}+BCB^{\mathsf{T}},
\end{eqnarray}
and so on: writing $(\mathbb{D}_\ell)_{ij}$ as a sum over sequences of $\ell-1$ intermediate states, the difference consists of the sequences that leave the retained space. For $\ell=2$ it is $e^2\Delta\mathbb{M}^2$ and is bounded by the TRK sum rule (Eq.~\eqref{eq:Delta_trk_bound}); for $\ell\ge3$ it involves $C$, which contains the large dipoles of the pseudocontinuum states. Neither choice is fully consistent: $\mathbb{D}_\ell$ matches the field part of the fDSE Hamiltonian but is not the image of any operator of the truncated space, whereas $\mathbb{M}^\ell$ derives from the unitary $\hat{U}_{\text{PZW}}$ restricted to that space, but the fDSE Hamiltonian is not its transform, which is the known noncommutativity of electronic truncation and gauge transformation~\cite{DiStefano2019}. Only $\mathbb{M}^\ell$ defines a free-photon number distribution $p_n^{\text{phys}}$. With $\mathbb{M}$, $\hat{n}_{\text{PZW}}$ has an integer spectrum and is the photon number of the tDSE; replacing $\mathbb{M}^2$ by $\mathbb{D}$, as in the fDSE Hamiltonian (Eq.~\eqref{eq:fDSE}), and the powers $\mathbb{M}^3$ and $\mathbb{M}^4$ by the matrix elements of $\hat{\mu}_\epsilon^3$ and $\hat{\mu}_\epsilon^4$, changes $\langle\hat{n}_{\text{PZW}}\rangle$ by at most 1\% for $g\le0.08$~a.u.\ and 3\% for $g=0.12$~a.u., and $\mathcal{G}^{(2)}_{\text{PZW}}(0)$, evaluated with all $n$, by less than 1\% for $g\le0.06$~a.u.\ and by at most 3\% for $g=0.12$~a.u.\ after the first 0.5~fs (less than 2\% after 2~fs). The largest difference occurs at $t=0$, where $\mathcal{G}^{(2)}_{\text{PZW}}(0)=\langle\hat{\mu}_\epsilon^4\rangle/\langle\hat{\mu}_\epsilon^2\rangle^2$ is purely electronic and the two choices give 2.205 and 2.347 for all $g$. The values of $\langle\hat{a}_{\text{PZW}}\rangle$ and $\langle\hat{n}_{\text{PZW}}\rangle$ shown in Figs.~\ref{fig:photon} and \ref{fig:averages} are expectation values with $\mathbb{D}$.

\paragraph{Second-order coherence.} In \textsf{enpd}, $\mathcal{G}^{(2)}_{\text{PZW}}(0)$ is the expectation value of $\hat{a}_{\text{PZW}}^{\dagger\,2}\hat{a}_{\text{PZW}}^{2}$ with the exact dipole operator $\hat{\mu}_\epsilon=\mathcal{Q}_{\text{mid}}R-|e|\hat{z}$ (Sec.~\ref{sec:methods_full}), so that the choice between $\mathbb{D}_\ell$ and $\mathbb{M}^\ell$ does not arise. In \textsf{npd--fDSE--MLS}, it is evaluated from $p_n^{\text{phys}}$, with the numerical tail described in Sec.~\ref{sec:res_statistics} replaced by a geometric continuation of the physical decay. The distribution has two regimes: for the first few photons the populations drop by more than a factor of three per photon (before 18~fs, $p_3^{\text{phys}}/p_2^{\text{phys}}$ and $p_4^{\text{phys}}/p_3^{\text{phys}}$ are below 0.3 at 98\% and 99.9\% of the instants), whereas in the tail consecutive populations are comparable (94\% of the ratios exceed 0.5). The elbow $n_c$ that separates them is the first $n\ge3$ with $p_{n+1}^{\text{phys}}/p_n^{\text{phys}}>0.3$. Populations up to $n_c$ are kept, and for $n>n_c$ they are replaced by $p_{n_c}^{\text{phys}}q^{\,n-n_c}$, which continues the decay with the last physical ratio, $q=\min(p_{n_c}^{\text{phys}}/p_{n_c-1}^{\text{phys}},\,0.5)$. The factorial moments $\Sigma_j=\sum_nn^{\underline{j}}\,p_n$ ($j=0,1,2$) of this distribution are then geometric series, and
\begin{eqnarray}\label{eq:G2geo}
 \mathcal{G}^{(2)}_{\text{PZW,geo}}(0)&=&\frac{\Sigma_2\Sigma_0}{\Sigma_1^2},\\
 \Sigma_j&=&\sum_{n=0}^{n_c}n^{\underline{j}}\,p_n^{\text{phys}}+p_{n_c}^{\text{phys}}\,\tau_j,\nonumber\\
 \tau_j&=&\sum_{i=0}^{j}\frac{j!}{i!}\,(n_c+1)^{\underline{i}}\,x^{\,j-i+1},\quad x=\frac{q}{1-q},\nonumber
\end{eqnarray}
with $n^{\underline{i}}=n(n-1)\cdots(n-i+1)$. Since $\Sigma_0$ is the norm, $\Sigma_1/\Sigma_0=\langle\hat{n}_{\text{PZW}}\rangle$, and $\Sigma_2/\Sigma_0=\langle\hat{n}_{\text{PZW}}(\hat{n}_{\text{PZW}}-1)\rangle$, the norm cancels in Eq.~\eqref{eq:G2geo}, and $q\to0$ recovers the cutoff at $n_c$. The result does not depend on the precise threshold: any value between 0.3 and 0.5 gives the same $n_c$ except at isolated instants (band in Fig.~\ref{fig:photon}(e)). It also agrees within 4\% with the distribution truncated at $n\le4$, whose populations are converged with respect to the integrator tolerances (Sec.~\ref{sec:res_statistics}). The comparison with \textsf{enpd} shows that the tail is an artifact of the ten-state calculation. Its most likely origins are an incomplete convergence of the photon-dependent part of the wave function, consistent with the sensitivity of the tail to the integrator tolerances and with the last natural population of $Q$ ($10^{-5}$ against $2\times10^{-7}$ in \textsf{enpd}), or spurious correlations with the pseudocontinuum states of the truncated fDSE Hamiltonian.

\paragraph{Time filter of the entropies.} The entropies of Sec.~\ref{sec:observables} oscillate with the rotation period of $\langle\hat{a}_{\text{PZW}}\rangle$. To show their slow evolution, each curve $y(t_k)$ is replaced by the local average $\tilde{y}(t_k)=\sum_jK_{jk}\,y(t_j)/\sum_jK_{jk}$, with the Gaussian kernel $K_{jk}=e^{-(t_j-t_k)^2/2\sigma^2}$, $\sigma=0.15$~fs, truncated at $|t_j-t_k|\le4\sigma$. This filter multiplies a component of frequency $\omega$ by $e^{-\omega^2\sigma^2/2}$, which removes more than 99\% of the signal for $\omega>0.76\,\omega_0$, and has a temporal resolution (full width at half maximum) of 0.35~fs; it therefore does not follow the initial rise of the entropies for $t\lesssim0.3$~fs. A rectangular average over one photon period, $2\pi/\omega_0$, would instead keep 15\% of the amplitude at $0.87\,\omega_0$, the rotation frequency of $\langle\hat{a}_{\text{PZW}}\rangle$ for $g=0.12$~a.u.